\documentclass[11pt,a4paper]{article}
\DeclareMathAlphabet{\scr}{U}{rsfs}{m}{n}

\usepackage{latexsym}
\usepackage{epsfig}
\usepackage[mathscr]{eucal}
\usepackage{amsfonts}
\usepackage{amscd}
\usepackage{cite}
\usepackage{array}   
\usepackage{amssymb}
\usepackage{colordvi}
\usepackage[centertags]{amsmath}
\usepackage{enumerate}
\usepackage{graphicx}
\usepackage{booktabs}
\usepackage{theorem}
\usepackage[footnotesize]{caption}
 \usepackage{slashed}
\usepackage{color}
\usepackage{ulem}
\newcommand{\newc}{\newcommand}
\newc{\be}{\begin{equation}}
\newc{\ee}{\end{equation}}
\newc{\bea}{\begin{eqnarray}}
\newc{\eea}{\end{eqnarray}}
\newc{\ol}{\overline}
\newc{\wt}{\widetilde}
\newc{\bs}{\boldsymbol}
\newc{\m}{\mathcal}
\newc{\la}{\langle}
\newc{\ra}{\rangle}
\newc{\pa}{\partial}

\newcommand{\non}{\nonumber} 
\newcommand{\beq}{\begin{eqnarray}} 
\newcommand{\eeq}{\end{eqnarray}} 
\newcommand{\bpmatrix}{\begin{pmatrix}}
\newcommand{\epmatrix}{\end{pmatrix}}
\newcommand{\ba}{\begin{array}}
\newcommand{\ea}{\end{array}}

\newcommand{\fr}{\frac}
\newcommand{\hc}{\text{ h.c.}}

\newcommand{\al}{\alpha}

\renewcommand{\ol}{\text{1l}}

\newcommand{\order}{\mathcal{O}(\alpha_t\alpha_s)}
\newcommand{\deltatwo}{\delta^{ \text{\tiny(}\!\text{\tiny2}\!\text{\tiny)}}\!}
\newcommand{\deltaone}{\delta^{ \text{\tiny(}\!\text{\tiny1}\!\text{\tiny)}}\!}
\newcommand{\Deltatwo}{\Delta^{ \text{\tiny(}\!\text{\tiny2}\!\text{\tiny)}}\!}
\newcommand{\Deltaone}{\Delta^{ \text{\tiny(}\!\text{\tiny1}\!\text{\tiny)}}\!}

\newcommand{\mueff}{\mu_{\text{eff}}}
\newcommand{\figref}[1]{Fig.~\ref{#1}}
\renewcommand{\eqref}[1]{Eq.~(\ref{#1})}

\newcommand{\appen}[1]{Appendix~\ref{#1}}

\newcommand{\DRb}{\overline{\text{DR}}}

\newcommand{\OS}{\text{OS}}
\newcommand{\MSb}{\overline{\text{MS}}}

\newcommand{\ie}{{\it i.e.\;}}

\newcommand{\bc}{\begin{center}}
\newcommand{\ec}{\end{center}}

\newcommand{\gev}{~\text{GeV}}
\newcommand{\mev}{~\text{MeV}}

\newcommand{\deltal}{\delta^{ \text{(}l\text{)}}\!}
\newcommand{\asla}{|\lambda|}
\newcommand{\aska}{|\kappa|}
\newcommand{\asbe}{s_\beta}
\newcommand{\acbe}{c_\beta}
\newcommand{\acphi}{c_{\phi_y}}
\newcommand{\asphi}{s_{\phi_y}}
\newcommand{\delzhu}{\deltal Z_{h_u}}
\newcommand{\deltalv}{\deltal v}
\newcommand{\deltalTB}{\deltal \tan\beta}
\newcommand{\deltalla}{\deltal \asla}
\newcommand{\deltalthd}{\deltal t_{h_d}}
\newcommand{\deltalthu}{\deltal t_{h_u}}
\newcommand{\deltaltad}{\deltal t_{a_d}}

\newcommand{\dMHpsq}{\deltal M^2_{H^\pm}}
\newcommand{\aMHp}{ M^2_{H^\pm}}
\newcommand{\Del}{\Delta^{(l)}\lambda^{\text{CT}}}

\newcommand{\s}{\newline \vspace*{-3.5mm}}

\begin{document}
\title{
\vspace*{-3cm}
\phantom{h} \hfill\mbox{\small KA-TP-14-2015}
\\[2cm]
\textbf{The Order $\order$ Corrections to the Trilinear Higgs Self-Couplings in the Complex NMSSM\\[4mm]}}

\date{}
\author{
Margarete M\"{u}hlleitner$^{1\,}$\footnote{E-mail:
  \texttt{margarete.muehlleitner@kit.edu}},
Dao Thi Nhung$^{2}$\footnote{E-mail: \texttt{thi.dao@kit.edu}},
Hanna Ziesche$^{1\,}$\footnote{E-mail: \texttt{hanna.ziesche@kit.edu}}
\\[9mm]
{\small\it
$^1$Institute for Theoretical Physics, Karlsruhe Institute of Technology,} \\
{\small\it Wolfgang-Gaede-Str. 1, 76131 Karlsruhe, Germany.}\\[3mm]
{\small\it
$^2$Institute of Physics, Vietnam Academy of Science and Technology, }\\
{\small\it 10 DaoTan, BaDinh, Hanoi, Vietnam.}\\[3mm]
}
\maketitle

\begin{abstract}
\noindent
A consistent interpretation of the Higgs data requires the same
precision in the Higgs boson masses and in the trilinear Higgs
self-couplings, which are related through their common origin from the
Higgs potential. In this work we provide the two-loop corrections at
order ${\cal O}(\alpha_t \alpha_s)$ in the approximation of
vanishing external momenta to the trilinear Higgs
self-couplings in the CP-violating Next-to-Minimal Supersymmetric
extension of the Standard Model (NMSSM). In the top/stop sector two different
renormalization schemes have been implemented, the OS and the $\DRb$
scheme. The two-loop corrections to the self-couplings are of the
order of 10\% in the investigated scenarios. The theoretical error, estimated
both from the variation of the renormalization scale and from the change of
the top/stop sector renormalization scheme, has been shown to be
reduced due to the inclusion of the two-loop corrections. 
\end{abstract}
\thispagestyle{empty}
\vfill
\newpage

\section{Introduction}
While the discovery of the Higgs boson by the LHC experiments ATLAS
\cite{Aad:2012tfa} and CMS \cite{Chatrchyan:2012ufa} certainly
marked a milestone for particle physics, it also triggered a change of
paradigm: The Higgs particle, formerly target of experimental
research, has become a tool in the quest for our understanding of
nature. Although the Standard Model (SM) of particle physics has been
tested at the quantum level and the discovered scalar particle behaves
SM-like \cite{Gouzevitch:2014pua} there are experimental and
theoretical arguments to assume it 
to be a low-energy effective theory of a more fundamental theory
appearing at some high scale. In the absence of any direct observation
of new states the study of the Higgs boson and its properties may
reveal the existence of beyond the SM (BSM) physics. In particular,
the discovered particle could be the SM-like Higgs boson of the
enlarged Higgs sector of a supersymmetric extension of the
SM. Supersymmetric (SUSY) theories
\cite{Volkov:1973ix,Wess:1974tw,Fayet:1976et,Fayet:1977yc,Fayet:1979sa,Farrar:1978xj,Dimopoulos:1981zb,Sakai:1981gr,Witten:1981nf,Nilles:1983ge,Haber:1984rc,Sohnius:1985qm,Gunion:1984yn,Gunion:1986nh,Lahanas:1986uc}
require the introduction of at least two complex Higgs doublets in
order to give masses to up- and down-type quarks and ensure an
anomaly-free theory. This minimal setup is extended by a complex
singlet superfield in the Next-to-Minimal Supersymmetric extension of
the SM (NMSSM) \cite{Fayet:1974pd,Barbieri:1982eh,Dine:1981rt,Nilles:1982dy,Frere:1983ag,Derendinger:1983bz,Ellis:1988er,Drees:1988fc,Ellwanger:1993xa,Ellwanger:1995ru,Ellwanger:1996gw,Elliott:1994ht,King:1995vk,Franke:1995tc,Maniatis:2009re,Ellwanger:2009dp}. After electroweak symmetry
breaking (EWSB) the NMSSM Higgs sector features seven Higgs bosons, which in
the CP-conserving case are three neutral CP-even, two neutral CP-odd
and two charged Higgs bosons. In contrast to the Minimal
Supersymmetric extension (MSSM)
\cite{Gunion:1989we,Martin:1997ns,Dawson:1997tz,Djouadi:2005gj}  
in the NMSSM CP-violation can occur in
the Higgs sector already at tree level. The additional sources of
CP-violation in SUSY theories are interesting not only because they 
clearly mark physics beyond the SM, but also because CP-violation
is an important ingredient for successful baryogenesis
\cite{Sakharov:1967dj}. From a 
phenomenological point of view it entails a plethora of interesting
new physics (NP) scenarios not excluded by experiment yet. \s

In order to study NP extensions, to properly interpret the
experimental data and to be able to distinguish different BSM
realizations, from the theory side we need as precise predictions as
possible not only for experimental observables\footnote{Neutral NMSSM
  Higgs production through gluon fusion and bottom-quark annihilation
  including higher order corrections has been discussed in \cite{Liebler:2015bka}.}
but also for the parameters of 
the theory under investigation. In the Higgs sector these are in
particular the Higgs boson masses and couplings. In the recent years
there has been quite some progress in the computation of the higher
order corrections to the Higgs boson masses of both the CP-conserving
and CP-violating NMSSM. Thus in the CP-conserving NMSSM after the
computation of the leading one-loop (s)top and (s)bottom contributions 
\cite{Ellwanger:1993hn,Elliott:1993ex,Elliott:1993uc,Elliott:1993bs,Pandita:1993tg}
and the chargino, neutralino as well as scalar one-loop contributions at
leading logarithmic accuracy \cite{Ellwanger:2005fh}, the full
one-loop contributions in the $\DRb$ renormalization scheme have first
been provided in~\cite{Degrassi:2009yq} and subsequently
in~\cite{Staub:2010ty}. In \cite{Degrassi:2009yq} also the order ${\cal O}
(\alpha_t \alpha_s + \alpha_b \alpha_s)$ corrections in the approximation of zero
external momenta have been given, and recently, first corrections
beyond order ${\cal O}(\alpha_t \alpha_s + \alpha_b \alpha_s)$ have
been published in \cite{Goodsell:2014pla,Goodsell:2015ira}. Our group
has calculated 
the full one-loop corrections in the Feynman diagrammatic approach in
a mixed $\DRb$-on-shell and in a pure on-shell renormalization scheme
\cite{Ender:2011qh}.  
In the CP-violating NMSSM the contributions to the mass corrections
from the third generation squark sector, from the charged particle
loops and from gauge boson contributions have been  
computed in the effective potential approach at one loop-level in 
Refs.~\cite{Ham:2001kf,Ham:2001wt,Ham:2003jf,Funakubo:2004ka,Ham:2007mt}.
The full one-loop and logarithmically enhanced two-loop effects in the
renormalization group approach have subsequently been given 
\cite{Cheung:2010ba}. We have contributed with the calculation of the 
full one-loop corrections in the Feynman diagrammatic approach
\cite{Graf:2012hh} and recently provided the two-loop corrections to
the neutral NMSSM Higgs boson masses in the Feynman diagrammatic
approach for zero external momenta at the order ${\cal O} (\al_t
\al_s)$ based on a mixed $\DRb$-on-shell renormalization scheme
\cite{Muhlleitner:2014vsa}.  
Several codes have been published for the evaluation of the NMSSM mass
spectrum from a user-defined input at a user-defined scale. The
Fortran package {\tt NMSSMTools}
\cite{Ellwanger:2004xm,Ellwanger:2005dv,Ellwanger:2006rn} computes 
the masses and decay widths in the 
CP-conserving $\mathbb{Z}^3$-invariant NMSSM and can be interfaced
with {\tt SOFTSUSY} 
\cite{Allanach:2001kg,Allanach:2013kza}, which provides 
the mass spectrum for a CP-conserving NMSSM,
also including the possibility of $\mathbb{Z}^3$ violation. Recently,
it has been extended to include also the CP-violating NMSSM
\cite{Domingo:2015qaa}. 
The spectrum of different SUSY models, 
including the NMSSM, can be generated by interfacing {\tt SPheno}
\cite{Porod:2003um,Porod:2011nf} with {\tt SARAH}
\cite{Staub:2010jh,Staub:2012pb,Staub:2013tta,Goodsell:2014bna,Goodsell:2014pla}. 
This is also the case for the recently published package {\tt FlexibleSUSY}
\cite{Athron:2014yba,Athron:2014wta}, when interfaced with {\tt
  SARAH}. All these codes include the Higgs mass corrections up to
two-loop order, obtained in the effective potential approach. 
The program package {\tt NMSSMCALC}
\cite{Baglio:2013vya,Baglio:2013iia} on the other hand, which
calculates the NMSSM Higgs masses and decay widths in the
CP-conserving and CP-violating NMSSM, provides the one-loop
corrections and the ${\cal O} (\alpha_t \alpha_s)$ corrections 
in the full Feynman diagrammatic approach, where the latter are
obtained in the approximation of vanishing external momenta. \s 

The Higgs self-couplings are intimately related to the Higgs boson
masses via the Higgs potential. For a consistent description therefore
not only the Higgs boson masses have to be provided at highest
possible precision, but also the Higgs self-couplings need to be
evaluated at the same level of accuracy. The trilinear Higgs
self-coupling enters the Higgs-to-Higgs decay widths. These can become
sizable in NMSSM Higgs sectors with light Higgs states in the
spectrum \cite{Ellwanger:2013ova,Munir:2013dya,King:2014xwa}, and via the total width these decays sensitively alter the
branching ratios of these states. Also Higgs pair production processes
are affected by the size of the trilinear Higgs self-couplings
\cite{Nhung:2013lpa,Han:2013sga,Wu:2015nba}. Their
determination marks a further step in our understanding of the Higgs
sector of EWSB
\cite{Djouadi:1999gv,Djouadi:1999rca,Muhlleitner:2000jj}. We have
provided the one-loop corrections to the trilinear Higgs
self-couplings for the CP-conserving NMSSM \cite{Nhung:2013lpa}. They 
have been calculated in the Feynman diagrammatic approach with
non-vanishing external momenta. The renormalization scheme that has
been applied is a mixture of On-Shell (OS) and $\overline{\mbox{DR}}$
conditions. In this paper we present, in the framework of the
CP-violating NMSSM, our computation of the dominant two-loop
corrections due to top/stop loops to the trilinear Higgs self-couplings of the
neutral NMSSM Higgs bosons. In addition, we give explicit formulae for
the leading one-loop corrections at order ${\cal O}(\alpha_t)$.  
We use the Feynman diagrammatic approach in the approximation of 
zero external momenta and furthermore work in the gaugeless limit. We find
that the determination of the two-loop corrections reduces the error on the
trilinear Higgs self-coupling due to unknown higher order corrections and
hence contributes to the effort of providing precise predictions for NMSSM
parameters and hence observables. We have furthermore expanded for
this paper the full one-loop corrections with full momentum dependence
to include  CP-violating effects. \s 

The outline of the paper is as follows. In
section~\ref{sec:outline-calculation} we set our notation, introduce
the NMSSM Higgs sector and present the determination of the loop-corrected
effective trilinear Higgs self-couplings. Section~\ref{sec:numerical}
is then dedicated to the numerical analysis. We discuss the effects of the loop
corrections on the trilinear Higgs self-couplings and the implications
for Higgs-to-Higgs decays. Section~\ref{sec:conclusion} contains our
conclusions. 

\section{The effective trilinear Higgs self-couplings in the NMSSM}
\label{sec:outline-calculation}
In this section we present the details of the calculation of the
effective trilinear Higgs self-couplings at order ${\cal
  O}(\alpha_t)$ and at order $\order$. We
closely follow the convention and notation of our paper on the Higgs
mass corrections in the complex NMSSM at order $\order$
\cite{Muhlleitner:2014vsa}. We therefore repeat here only the most
important definitions relevant for our calculation. We work in the
framework of the complex NMSSM with a scale invariant superpotential
and a discrete $\mathbb{Z}^3$ symmetry. In terms of the two Higgs
doublets  $H_d$ and $H_u$, and the scalar singlet $S$, the Higgs potential reads,
\beq
V_{H}  &=& (|\lambda S|^2 + m_{H_d}^2)H_{d,i}^* H_{d,i}+ (|\lambda S|^2
+ m_{H_u}^2)H_{u,i}^* H_{u,i} +m_S^2 |S|^2 \nonumber \\
&& + \fr18 (g_2^2+g_1^{2})(H_{d,i}^* H_{d,i}-H_{u,i}^* H_{u,i} )^2
+\fr12g_2^2|H_{d,i}^* H_{u,i}|^2 \label{eq:higgspotential} \\ 
&&   + |-\epsilon^{ij} \lambda  H_{d,i}  H_{u,j} + \kappa S^2 |^2+
\big[-\epsilon^{ij}\lambda A_\lambda S   H_{d,i}  H_{u,j}  +\fr13 \kappa
A_{\kappa} S^3+\hc \big] \,.
\nonumber
\eeq
The indices of the fundamental representation of $SU(2)_L$  are denoted by
$i,j=1,2$, and $\epsilon_{ij}$ is the totally antisymmetric tensor
with $\epsilon_{12}= \epsilon^{12} = 1$.  The dimensionless parameters
$\lambda$ and $\kappa$ and the soft SUSY breaking trilinear couplings
$A_\lambda$ and $A_\kappa$ can in general be complex.
The $U(1)_Y$ and $SU(2)_L$ gauge couplings are given by $g_1$ and
$g_2$, respectively. In order to obtain the Higgs boson masses,
trilinear and quartic Higgs self-couplings from the Higgs potential,
the Higgs doublets and the singlet field are replaced by the expansions about their
vacuum expectation values (VEVs), $v_d, v_u$ and $v_s$,  
\beq
H_d =
 \bpmatrix \fr{1}{\sqrt 2}(v_d + h_d +i a_d)\\ h_d^- \epmatrix,\quad
H_u = e^{i\varphi_u}\bpmatrix
h_u^+ \\ \fr{1}{\sqrt 2}(v_u + h_u +i a_u)\epmatrix,\quad
S= \fr{e^{i\varphi_s}}{\sqrt 2}(v_s + h_s +ia_s),~
\label{eq:Higgs_decomposition} 
\eeq
where two additional phases,  $\varphi_u$ and $\varphi_s$, have been 
introduced. Note, that in order to keep the Yukawa coupling neutral we
absorb the phase $\varphi_u$ into the left- and right-handed top
fields, which of course affects all couplings involving only one top
quark \cite{Muhlleitner:2014vsa}. \s

We work in the approximation of zero external momenta and call the
thus derived self-couplings 'effective' self-couplings. The
(loop-corrected) self-couplings are automatically real in this approach. In the
interaction basis, the effective trilinear 
Higgs self-couplings at order $\order$ can be cast into the form 
\be
  \label{eq:effcoup}
  \Gamma_{\phi_i\phi_j\phi_k}=\lambda_{\phi_i\phi_j\phi_k}+\Deltaone\lambda_{\phi_i\phi_j\phi_k}
  +\Deltatwo\lambda_{\phi_i\phi_j\phi_k},
\ee
with $\phi=(h_d,h_u,h_s,a_d,a_u,a_s)$ and $i,j,k=1,\ldots,6$. The
first term represents the tree-level trilinear couplings, which can
directly be derived from the tree-level Higgs potential
\eqref{eq:higgspotential} by taking the derivative
\be 
\lambda_{\phi_i\phi_j\phi_k}= \frac{\pa^3 V_H}{\pa \phi_i \pa \phi_j \pa
  \phi_k} \;. 
\ee
Explicit expressions for these couplings can be found in 
\appen{sec:tree-leve-trilinear}. The second and third terms denote 
the one- and two-loop corrections to the Higgs self-couplings. 
They can be obtained by either taking the derivative of the
corresponding loop-corrected effective potential or by using the
Feynman diagrammatic approach in the approximation of zero external
momenta.  At one-loop level we use both methods and find that the
results obtained in these two different approaches
agree as expected. However, at two-loop level, for the sake of 
automatization of our codes we solely employ the Feynman diagrammatic
approach.\footnote{In the effective potential approach the derivatives
which are taken to get the Higgs self-couplings lead to very large
intermediate expressions, that are not practical to be used for automatization.}  
Therefore only the latter is described in the following. \s

In order to obtain the effective trilinear couplings in
the mass eigenstate basis, the self-couplings in the
interaction basis have to be rotated to the mass basis by applying the
rotation matrix ${\cal R}^{(l)}$. In detail, we
have,
\be
\Phi = (H_1,H_2,H_3,H_4,H_5,G)\,, \quad \Phi^T= {\cal R}^{(l)}\,
\phi^T \;, 
\label{eq:phirotfull}
\ee
where $l=1,2$ stands for the loop order and $\Phi$ for the loop-corrected
mass eigenstates. These are denoted by upper case $H$ and ordered by
ascending mass with $M_{H_1}\leq M_{H_2}\leq M_{H_3} \leq M_{H_4} \leq M_{H_5}$. The
neutral Goldstone boson $G$ has been singled out. Note in particular,
that the mass eigenstates are no CP eigenstates any more since we work
in the CP-violating NMSSM. In order to be as precise as possible in
the computation of the loop-corrected trilinear Higgs self-couplings
in the mass eigenstate basis, we employ the most precise rotation
matrix ${\cal R}^{(l)}$ that is available. This means, that we rotate to the
mass eigenstates $H_i$ ($i=1,...,5$) at two-loop order. 
The loop-corrected rotation matrix ${\cal R}^{(l)}$ is computed by the
Fortran package 
{\tt NMSSMCALC} \cite{Baglio:2013vya,Baglio:2013iia} where the zero
momentum approximation is employed, so 
that the matrix is unitary. In particular, the rotation matrix
includes the complete electroweak (EW) corrections at one-loop order and the
order ${\cal O}(\alpha_t \alpha_s)$ corrections at two-loop level. For
more details see
\cite{Ender:2011qh,Graf:2012hh,Muhlleitner:2014vsa}. \s

The rotation matrix ${\cal R}^{(l)}$ can be decomposed in the rotation
matrix ${\cal R}$, that rotates the interaction eigenstates to the tree-level
mass eigenstates $\Phi^{(0)} \equiv (h_1,h_2,h_3,h_4,h_5,G)$,
singling out the Goldstone boson $G$, and in the finite wave-function
renormalization factor ${\bf Z}$, {\it cf.}~\cite{Ender:2011qh,Nhung:2013lpa},
\beq
{\cal R}^{(l)}_{is} = {\bf Z}_{ij} {\cal R}_{js} \;, \quad
i,j,s=1,..,6 \;. \label{eq:mixmatz}
\eeq
With this definition, the loop-corrected effective trilinear couplings
between the Higgs bosons in the 2-loop 
mass eigenstate basis are hence given by $(i,i',j,j',k,k'=1,...,5)$
\be 
\Gamma_{\Phi_i\Phi_j\Phi_k}= {\bf Z}_{ii^\prime}{\bf
  Z}_{jj^\prime}{\bf Z}_{kk^\prime}
\Gamma_{h_{i^\prime}h_{j^\prime}h_{k^\prime}}
\;, \label{eq:mixmat} 
\ee
where the couplings in the tree-level mass eigenstates
$\Gamma_{h_{i^\prime}h_{j^\prime}h_{k^\prime}}$ are obtained from
Eq.~(\ref{eq:effcoup}) by rotation with ${\cal R}$. 

\subsection{The order ${\cal O}(\al_t)$
  corrections\label{ssect:one-loop}}
In this subsection we present the one-loop corrections at order
${\cal O}(\al_t)$. Due to the large top quark Yukawa coupling, at
one-loop level the corrections from the top/stop sector are the
dominant corrections to the Higgs boson masses and self-couplings. This is in
particular true for the SM-like Higgs boson. The latter must be dominantly
$h_u$-like, inducing via the top loop a sufficiently large coupling to
the gluons, so that its rates are in accordance with the measured 
signal rates of the discovered Higgs boson, which at the LHC is
dominantly produced through gluon fusion. The restriction to the order
${\cal O}(\al_t)$ corrections with large top/stop masses in the loops
furthermore ensures the approximation of zero external momenta to
be reliable. This approximation breaks down if the masses of the
particles running in the loops are small.\footnote{Scenarios with
  light Higgs bosons are mostly 
precluded as otherwise the kinematically allowed Higgs-to-Higgs decays
would lower the branching ratios of the SM-like Higgs boson into the
other SM particles to values not compatible with the experimental data
any more. However, other light particles running in the loops could
spoil the validity of the zero momentum approximation.}
For the numerical analysis presented in section~\ref{sec:numerical} we took care
to choose scenarios where all possibly involved loop particles 
are sufficiently heavy so that not only the approximation of zero
external momenta works well but also the order ${\cal O}(\alpha)$
corrections do not play a significant role. The full EW 
and the order ${\cal O}(\alpha_t)$ corrections differ by less than 4\%
for the chosen scenarios as we explicitly verified. \s

In the following we give the analytic formulae for the one-loop
order ${\cal O}(\alpha_t)$ corrections to the trilinear Higgs
self-couplings in the interaction basis at vanishing external
momenta. These formulae are compact enough to be easily implemented in
computer codes.  
In order to extract only the $\mathcal{O}(\al_t)$  and later on the 
$\mathcal{O}(\al_t\al_s)$ corrections we neglect all $D$-term
contributions to the Higgs potential and to the stop mixing matrix \ie we
work in the gaugeless limit, where the electric coupling $e$ and the $W$ 
and $Z$ boson masses $M_W$ and $M_Z$ are taken to be zero but the vacuum
expectation value $v$ and the weak mixing angle $\theta_W$ are
kept finite. In this approximation, the stop mass matrix reads  
\begin{equation}
\mathcal{M}_{\tilde t}=
\bpmatrix
 m_{\tilde{Q}_3}^2+m_t^2 & m_t \left(A_t^*
   e^{-i\varphi_u}-\fr{\mueff}{\tan\beta}\right) \\[2mm] 
m_t \left(A_t e^{i\varphi_u}- \fr{\mueff^*}{\tan\beta}\right) & m_{\tilde{t}_R}^2+m_t^2
\epmatrix\,, 
\end{equation}
where $m_t$ denotes the top quark mass and the effective higgsino
mixing parameter   
\beq
\mueff= \fr{\lambda v_s e^{i\varphi_s}}{\sqrt{2}} \label{eq:mueff}
\eeq
and the ratio of the two VEVs $v_u$ and $v_d$, 
\beq
\tan\beta= \frac{v_u}{v_d} \;,
\eeq
have been introduced. The soft SUSY breaking masses
$m_{\tilde{Q}_3}$ and $m_{\tilde{t}_R} $ are real, whereas the 
trilinear coupling $A_t \equiv |A_t| \exp(i\varphi_{A_t})$ is in general complex. The
matrix is diagonalized by a unitary matrix 
$\mathcal{U}_{\tilde t}$, rotating the interaction states $\tilde{t}_L$
and $\tilde{t}_R$ to the mass eigenstates $\tilde{t}_1$ and
$\tilde{t}_2$, 
\bea
  (\tilde t_1,\tilde t_2)^T &=& \mathcal{U}_{\tilde t}~(\tilde t_L,\tilde t_R)^T\,\\
 \text{diag}(m_{\tilde t_1}^2,m_{\tilde t_2}^2)&=&\mathcal{U}_{\tilde t}~\mathcal{M}_{\tilde t}~\mathcal{U}_{\tilde t}^\dagger\,.
\eea

The order $\mathcal{O}(\al_t)$ corrections to the trilinear  Higgs
self-couplings in the interaction basis are decomposed as 
\be 
\Deltaone\lambda_{\phi_i\phi_j\phi_k} =
\Deltaone\lambda^{\text{UR}}_{\phi_i\phi_j\phi_k} +
\Deltaone\lambda^{\text{CT}}_{\phi_i\phi_j\phi_k} \; . \label{eq:decomposition}
\ee
The first term denotes the unrenormalized part arising from the
one-loop diagrams with tops and stops running in the loops. The
explicit expressions for 
$\Deltaone\lambda^{\text{UR}}_{\phi_i\phi_j\phi_k} $ are given in
\appen{sec:one-loop-trilinear}.  The contributions from the parameter
counterterms are collected in the second part
$\Deltaone\lambda^{\text{CT}}_{\phi_i\phi_j\phi_k}$. Their explicit
expressions in terms of the counterterms, defined in the following,
are given in \appen{sec:trilinearCT}.\s

For the order $\mathcal{O}(\al_t)$  and and the order $\mathcal{O}(\al_t \al_s)$
corrections, we need to renormalize  the following set of parameters
\cite{Muhlleitner:2014vsa},\footnote{As we work in the gaugeless
  limit, {\it i.e.}~$e=0$ and $M_W=M_Z=0$ but $v\ne 0$ and
  $\sin\theta_W \ne 0$, it is convenient to choose $v$ and $\sin\theta_W$ in the
  computation of the higher order corrections, instead of $M_W$ and
  $M_Z$. Note that $\sin\theta_W$ does not appear in the Higgs
  potential in the gaugeless limit. See also
  \cite{Muhlleitner:2014vsa}, for more details.} 
\be 
t_{h_d},t_{h_u},t_{h_s},t_{a_d},t_{a_s},M_{H^\pm}^2,v,\tan\beta,|\lambda|\,,
\ee
where $t_{\phi}$, $\phi=h_d,h_u,h_s,a_d,a_s$ denote the five
independent tadpoles, $M_{H^\pm}$ stands for the mass of the charged
Higgs boson and $v \approx 246$~GeV is given by
\be  
v^2=v_u^2+v_d^2 \;. 
\ee
In order to renormalize the parameters, they are replaced by the
renormalized ones and the corresponding counterterms as follows:
\begin{align} 
t_{\phi} &\rightarrow t_{\phi} +\deltaone t_{\phi}  +\deltatwo 
t_{\phi} \qquad\qquad\quad\quad \text{with}~~\phi=h_d,h_u,h_s,a_d,a_s
\;,\label{eq:tphi}\\
M_{H^\pm}^2 &\rightarrow M_{H^\pm}^2+  \deltaone M_{H^\pm}^2+  \deltatwo M_{H^\pm}^2\;,\\
v &\rightarrow v + \deltaone v +  \deltatwo v\;,\\
\tan\beta &\rightarrow \tan\beta  +  \deltaone \tan\beta+  \deltatwo \tan\beta\;,\\
|\lambda| &\rightarrow |\lambda|+  \deltaone |\lambda| +  \deltatwo
|\lambda|\; .
\label{eq:sphiy}
\end{align}
Here the superscript $(1)$ denotes the counterterms of ${\cal O}(\al_t)$ 
and the superscript $(2)$ the counterterms of ${\cal O}(\al_t\al_s)$.\s

In addition to the parameter renormalization, also the wave function
renormalization of the Higgs fields is needed in order to obtain a UV
finite result. At ${\cal O}(\al_t)$ and ${\cal O}(\al_t\al_s)$, only
the Higgs doublet $H_u$ has a non-vanishing wave function
renormalization
counterterm~\cite{Muhlleitner:2014vsa}, which is
introduced as  
\be 
H_u\rightarrow \left(1+ \fr{1}2 \deltaone Z_{H_u}+ \fr{1}2 \deltatwo
  Z_{H_u} \right)H_u \; . 
\ee
The parameters are renormalized in a mixed $\mbox{OS}$-$\DRb$ renormalization
scheme as described in \cite{Muhlleitner:2014vsa}. In this scheme part
of the parameters, that are directly related to ``physical''
quantities, are renormalized on-shell, and the remaining parameters
are defined via $\DRb$ conditions, as 
\be
\underbrace{ t_{h_d},t_{h_u},t_{h_s},t_{a_d},t_{a_s},M_{H^\pm}^2,v}_{\mbox{on-shell
 scheme}},
\underbrace{\tan\beta,|\lambda|}_{\overline{\mbox{DR}} \mbox{ scheme}}
\;. 
\label{eq:mixedcond}
\ee
While it is debatable if the tadpole parameters can be called physical
quantities, their introduction is motivated by physical
interpretation, so that in slight abuse of the language we call their
renormalization conditions on-shell. 
For the wavefunction renormalization of the Higgs fields, the $\DRb$
scheme is employed. 
Note that this procedure is applied for both the order ${\cal
  O}(\al_t)$ and the order ${\cal O}(\al_t\al_s)$ corrections.
We do not repeat the renormalization conditions here, since they are
introduced in detail in \cite{Muhlleitner:2014vsa}. We give, however, the
explicit expressions for the counterterms of order ${\cal O}(\al_t)$.
For the OS renormalization constants at order ${\cal O}(\al_t)$ we
find in $D=4-2\epsilon$ dimensions:
\bea
\fr{\deltaone v}{v} &=&\fr{c_W^2}{2s_W^2}\left(\fr{\deltaone
    M_Z^2}{M_Z^2}-\fr{\deltaone M_W^2}{M_W^2}\right)+\fr12
\fr{\deltaone M_W^2}{M_W^2}  \\
\deltaone M_{H^\pm}^2&=&\frac{3 m_t^2c_{\beta}^2}{8 \pi^2 s_{\beta}^2 v^2}
\Big( \text{A}_0 (m_{\tilde Q_3}^2)-2 
\text{A}_0(m_t^2)+ |\mathcal{U}_{\tilde{t}_{12}}|^2
\text{A}_0(m_{\tilde{t}_1}^2) +
|\mathcal{U}_{\tilde{t}_{22}}|^2 \text{A}_0 (m_{\tilde{t}_2}^2)] \,
\nonumber \\
&+&\left| m_t |\mathcal{U}_{\tilde{t}_{11}}|+|A_t| 
  e^{i \phi_x} |\mathcal{U}_{\tilde{t}_{12}}|+\frac{|\lambda| t_{\beta}
    v_s |\mathcal{U}_{\tilde{t}_{12}}|}{\sqrt{2}}\right|^2
\text{B}_0(0,m_{\tilde Q_3}^2,m_{\tilde{t}_1}^2)
\nonumber \\
&+&\left| m_t |\mathcal{U}_{\tilde{t}_{21}}|+|A_t| 
  e^{i \phi_x} |\mathcal{U}_{\tilde{t}_{22}}|+\frac{|\lambda| t_{\beta}
    v_s |\mathcal{U}_{\tilde{t}_{22}}|}{\sqrt{2}}\right|^2
\text{B}_0(0,m_{\tilde Q_3}^2,m_{\tilde{t}_2}^2) \Big) \\
\delta^{\scriptscriptstyle{(1)}}t_{h_d}&=& \frac{3 |\lambda| m_t^2 v_s
  \left(\sqrt{2} c_{\beta} |\lambda| v_s-2 |A_t| s_{\beta}
    c_{\varphi_x}\right)}{16
  \sqrt{2}  \pi ^2 s_{\beta}^2 v}\left(\frac{1}{\epsilon} + F_1\right)
\eeq
\bea
\delta^{\scriptscriptstyle{(1)}}t_{h_u}&=&-\frac{3 m_t^2}{16   \pi ^2 s_{\beta}^2 v}\frac{1}{\epsilon} \left[
\sqrt{2} |A_t|  |\lambda| v_s c_{\beta} c_{\varphi_x}-2
    s_{\beta}\left(|A_t|^2 +m_{\tilde{t}_1}^2+m_{\tilde{t}_2}^2-2 m_t^2\right)
     \right] \nonumber \\
&-&\frac{3 m_t^2}{16  \pi ^2 s_{\beta}^2 v} \bigg[\sqrt{2}
  |A_t| |\lambda| v_s  c_{\beta} c_{\varphi_x}F_1-2
s_{\beta} \bigg(|A_t|^2 F_1+
  m_{\tilde{t}_1}^2+
  m_{\tilde{t}_2}^2-2  m_t^2\nonumber
\\ &-& m_{\tilde{t}_1}^2
    \log\frac{m_{\tilde{t}_1}^2}{Q^2}-
    m_{\tilde{t}_2}^2 \log\frac{m_{\tilde{t}_2}^2}{Q^2}+2 
    m_t^2 \log\frac{m_t^2}{Q^2}\bigg)\bigg] \\
\delta^{\scriptscriptstyle{(1)}}t_{h_s}&=&\frac{v c_\beta}{v_s}
\delta^{\scriptscriptstyle{(1)}}t_{h_d} \\
\delta^{\scriptscriptstyle{(1)}}t_{a_d}&=&\frac{3 |A_t| |\lambda| m_t^2 v_s
  s_{\varphi_x}}{8 \sqrt{2} \pi ^2 s_{\beta} v}\left(\frac{1}{\epsilon} +F_1\right) \\
\delta^{\scriptscriptstyle{(1)}}t_{a_s}&=&\frac{v c_\beta}{v_s} \delta^{\scriptscriptstyle{(1)}}t_{a_d}
\eea 
with 
\bea \fr{\deltaone
  M_W^2}{M_W^2}&=&-\fr{3m_t^2}{8\pi^2 v^2}\fr1\epsilon -\fr{3}{16 \pi^2
  v^2}\bigg[m_t^2 - 2m_t^2\log\fr{m_t^2}{Q^2}+|\mathcal{U}_{\tilde
  t_{11}}|^2 F_0(m_{\tilde t_1}^2,m_{\tilde Q_3}^2)\\ \nonumber &&+
|\mathcal{U}_{\tilde t_{21}}|^2 F_0(m_{\tilde
  t_2}^2,m_{\tilde Q_3}^2)\bigg] \\
\fr{\deltaone M_Z^2}{M_Z^2}&=& -\fr{3m_t^2}{8\pi^2
  v^2}\fr1\epsilon-\fr{3}{16 \pi^2 v^2}\bigg[- 2m_t^2
\log\fr{m_t^2}{Q^2} +|\mathcal{U}_{\tilde t_{11}}|^2|\mathcal{U}_{\tilde
  t_{12}}|^2 F_0(m_{\tilde t_1}^2,m_{\tilde t_2}^2)\bigg] \eeq and \bea
F_0(x,y) &=& x+y -\fr{2xy}{x-y}\log\fr{x}{y}\,,\\
F_1&=&\frac{m_{\tilde{t}_2}^2-m_{\tilde{t}_1}^2+m_{\tilde{t}_1}^2
    \log\frac{m_{\tilde{t}_1}^2}{Q^2}-m_{\tilde{t}_2}^2
    \log\frac{m_{\tilde{t}_2}^2}{Q^2}}{m_{\tilde{t}_2}^2-m_{\tilde{t}_1}^2}. \eea And for the $\DRb$
renormalization constants we get: \be \deltaone
Z_{H_u}=\fr{-3m_t^2}{8\pi^2 v^2 \sin^2\beta}\fr1\epsilon\,,\qquad
\deltaone \tan\beta= \fr12\tan\beta \, \deltaone Z_{H_u}\,,\qquad
\deltaone|\lambda|=\frac{-|\lambda|}{2}\deltaone Z_{H_u}\,. \ee 
Here 
$\phi_x=\varphi_u + \varphi_s+ \varphi_\lambda + \varphi_{A_t}$, with
$\varphi_\lambda$ and $\varphi_{A_t}$ being the complex phase of
$\lambda$, and accordingly of $A_t$, has been introduced. Furthermore we use
$c_\beta \equiv \cos \beta$ etc. The functions $
\text{A}_0\!\left(m^2\right)$ and $
\text{B}_0\!\left(p^2,m_{1}^2,m_{2}^2\right)$ denote
the scalar one-point and two-point functions, respectively, in the
convention of \cite{Hahn:looptools}, and $Q$ is the renormalization scale. 

\subsection{The order ${\cal O}(\al_t\al_s)$
  corrections \label{ssec:two-loop}} 
\begin{figure}[ht]
  \centering
\includegraphics[scale=0.6]{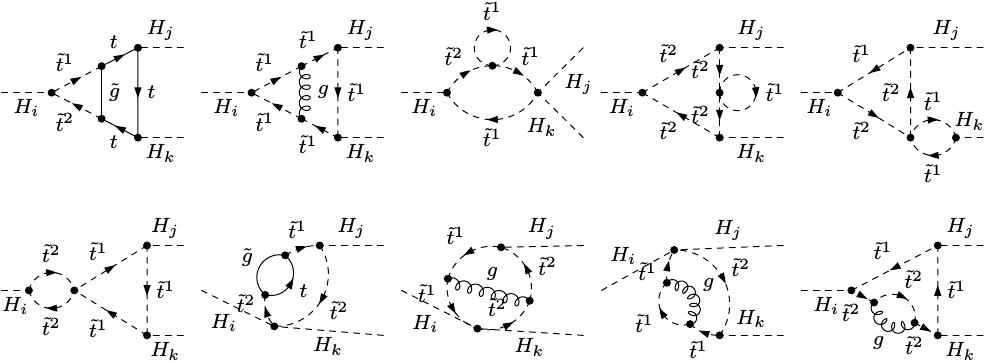}
  \caption{Sample of genuine two-loop diagrams contributing to the
    ${\cal O}(\al_t\al_s)$ corrections to the trilinear Higgs
    self-couplings between $H_i$, $H_j$ and $H_k$ ($i,j,k=1,...,5$).}
  \label{fig:2loop-diags}
\end{figure}

In order to obtain the order ${\cal O}(\al_t\al_s)$ corrections we use the Feynman
diagrammatic approach in the approximation of zero external
momenta. These corrections are composed of
\bea 
\Deltatwo\lambda_{\phi_i\phi_j\phi_k} =
\Deltatwo\lambda^{\text{UR}}_{\phi_i\phi_j\phi_k}+
\Deltatwo\lambda^{\text{CT1L}}_{\phi_i\phi_j\phi_k} +
\Deltatwo\lambda^{\text{CT2L}}_{\phi_i\phi_j\phi_k} \;.
\eea 
The first part consists of the contributions from genuine two-loop
diagrams. These must contain either a gluon or gluino or a four-stop 
coupling in order to give a contribution of order ${\cal O} (\alpha_t
\alpha_s)$. Some sample diagrams are presented in
\figref{fig:2loop-diags}.\footnote{Note that we work in the
  CP-violating NMSSM, so that we have trilinear couplings between all five
  neutral Higgs mass eigenstates.} In the approximation of zero
external momenta all 
two-loop three-point functions can be reduced to the product of two
one-loop tadpoles and to the two-loop one-point integral which are
presented analytically in the literature
\cite{Davydychev:1992mt,Ford:1992pn,Scharf:1993ds,Weiglein:1993hd,Berends:1994ed,Martin:2001vx,Martin:2005qm}. \s

\begin{figure}[h]
 \centering
\includegraphics[scale=0.6]{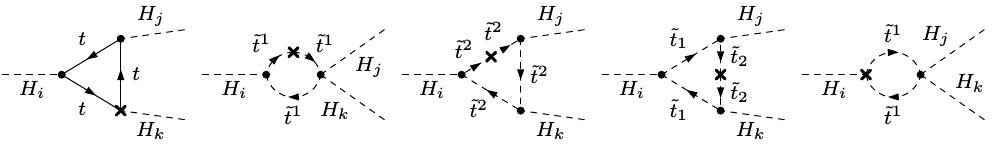}
 \caption{Some representative one-loop diagrams with one-loop
   counterterm insertion contributing to the  ${\cal O}(\al_t\al_s)$
   corrections to the trilinear Higgs self-couplings.}
 \label{fig:1loop-1lCT-diags}
\end{figure}
 The second part $\Deltatwo\lambda^{\text{CT1L}}_{\phi_i\phi_j\phi_k}$
denotes the contributions arising from the one-loop diagrams with top
quarks and stops as loop particles and with one insertion of a
counterterm of order ${\cal O}(\al_s)$ from the top/stop
sector. Some representative diagrams for this set are depicted in
\figref{fig:1loop-1lCT-diags}. 
The parameters of the top/stop and bottom/sbottom
sectors are renormalized at order ${\cal O}(\alpha_s)$. The bottom
quarks are treated as massless, so that the left- and right-handed
sbottom states do not mix and only the left-handed sbottom with a mass
of $m_{\tilde{Q}_3}$ contributes. We choose the set of independent
parameters entering the top/stop and bottom/sbottom sector, that we
renormalize, to be given by 
\beq
m_t \;, \quad m_{\tilde{Q}_3} \;, \quad m_{\tilde{t}_R} \quad
\mbox{and} \quad A_t \;.
\eeq
Note that  $ A_t$ is in general complex. We renormalize
these parameters both in the $\DRb$ and in the $\OS$ scheme. 
The definition of their counterterms can be found in
\cite{Muhlleitner:2014vsa}\footnote{Note that our OS scheme does not
  take into account terms proportional to $\epsilon$.}. According
to the SUSY Les Houches Accord (SLHA) 
\cite{Skands:2003cj,Allanach:2008qq} convention, 
$m_{\tilde{Q}_3}$, $m_{\tilde{t}_R}$ and $A_t$ are given as $\DRb$
parameters. When we choose the $\OS$ scheme these parameters need
finite shifts for the conversion into $\OS$ parameters. In the $\DRb$
scheme on the other hand, the given top pole mass must be translated into
a $\DRb$ mass. These translations according to our conventions are
described in detail in \cite{Muhlleitner:2014vsa}.  \s

The third part consists of contributions arising from the order ${\cal
  O}(\al_t\al_s)$ counterterms. The explicit expressions of
$\Deltatwo\lambda^{\text{CT2L}}_{\phi_i\phi_j\phi_k}$ in terms of 
$\deltatwo t_\phi$, 
$\deltatwo M_{H^\pm}^2,$ $\deltatwo v,$ $\deltatwo \tan\beta$,
$\deltatwo |\lambda|$ and $\deltatwo Z_{H_u}$ are the same as in the
one-loop case after 
replacing the one-loop by the two-loop counterterms. The formulae are
given in \appen{sec:trilinearCT}. For the exact definitions of
the two-loop counterterms, we refer the reader to
\cite{Muhlleitner:2014vsa}. \s 

Our results have been obtained in two independent calculations. For
the generation of the amplitudes we have employed \texttt{FeynArts}  
\cite{Kublbeck:1990xc,Hahn:2000kx} using in one calculation a model
file created by \texttt{SARAH}
\cite{Staub:2009bi,Staub:2010jh,Staub:2012pb,Staub:2013tta} and in the
other calculation a 
model file based on the one presented in \cite{Hahn:MSSM} which has
been extended by our group to the case of the NMSSM. The
contraction of the Dirac  matrices was done with 
\texttt{FeynCalc} \cite{Mertig:1990an}. The reduction to master
integrals was performed using the program \texttt{TARCER}
\cite{Mertig:1998vk}, which is based on a reduction algorithm
developed by Tarasov \cite{Tarasov:1996br,Tarasov:1997kx} and which is 
included in \texttt{FeynCalc}. We have applied dimensional reduction
\cite{Siegel:1979wq,Stockinger:2005gx} in the manipulation of the
Dirac algebra and in the tensor reduction. In our calculation no
$\gamma_5$ terms appear that require a special treatment in $D$
dimensions, so that we take $\gamma_5$ to be anti-symmetric with all
other  Dirac matrices. The cancellation of the single pole and double
poles has been checked. The results of the two computations are in
full agreement. We furthermore compared our results in the limit of
the real MSSM with Ref.~\cite{Brucherseifer:2013qva} where the two-loop ${\cal
  O}(\alpha_t \alpha_s)$ corrections to the MSSM Higgs self-couplings
were given, and we found agreement between the two computations.  

\section{Numerical analysis \label{sec:numerical}}
\subsection{Scenarios \label{ssec:constraints}}
For the numerical analysis of the impact of the higher order
corrections on the Higgs self-couplings we made sure to choose
scenarios that comply with the experimental constraints. In order to
find viable scenarios we performed a scan in the NMSSM parameter
space. We checked the scenarios for their accordance with the LHC Higgs
data by using the programs {\tt 
  HiggsBounds} \cite{Bechtle:2008jh,Bechtle:2011sb,Bechtle:2013wla}
and {\tt HiggsSignals} \cite{Bechtle:2013xfa}. The programs require
as inputs the effective couplings of the Higgs bosons, normalized to the
  corresponding SM values, as well as the masses, the widths and the
  branching ratios of the Higgs bosons. These have been obtained for
  the SM and NMSSM Higgs bosons from the Fortran code 
  {\tt NMSSMCALC} \cite{Baglio:2013vya,Baglio:2013iia}. A remark is in order for the 
  loop-induced Higgs couplings to gluons and photons. The effective
  NMSSM Higgs boson coupling to the gluons normalized to the
  corresponding 
  coupling of a SM Higgs boson with same mass is obtained by taking
  the ratio of the partial widths for the Higgs decays into gluons in
  the NMSSM and the SM, respectively. The QCD corrections up to 
next-to-next-to-next-to leading order in the limit of heavy quarks
\cite{Inami:1982xt,Djouadi:1991tka,Spira:1993bb,Spira:1995rr,Kramer:1996iq,Chetyrkin:1997iv,Chetyrkin:1997un,Schroder:2005hy,Chetyrkin:2005ia,Baikov:2006ch}
and squarks \cite{Dawson:1996xz,Djouadi:1996pb} are included. As
the EW corrections are unknown for the NMSSM Higgs boson
decays, they are consistently neglected also in 
the SM decay width. The loop-mediated effective Higgs coupling to the
photons has been obtained analogously. 
Here the NLO QCD corrections
to quark and squark loops including the full mass dependence for the quarks
\cite{Spira:1995rr,Zheng:1990qa,Djouadi:1990aj,Dawson:1992cy,Djouadi:1993ji,Melnikov:1993tj,Inoue:1994jq}
and squarks \cite{Muhlleitner:2006wx} are taken into 
account. The EW corrections, which are unknown for the
SUSY case, are neglected also in the SM. \s

For the numerical analysis of the corrections to the Higgs
self-couplings, we have chosen two parameter sets that 
fulfill the above constraints. For both scenarios we use the SM input
parameters \cite{Agashe:2014kda,Jegerlehner:2011mw}
\begin{align}
\alpha(M_Z)&=1/128.962 \,, &\alpha^{\overline{\mbox{MS}}}_s(M_Z)&=
0.1184 \,, &M_Z&=91.1876\, 
\gev\,, \\ \non
M_W&=80.385\,\gev \,,   &m_t&=173.5\,\gev \,,
&m^{\MSb}_b(m_b^{\MSb})&=4.18\,\gev \;. \label{eq:param1}
\end{align} 
In the numerical evaluation, however, we chose to use the running
$\al_s^{\DRb}$. It is obtained by converting the
$\alpha^{\overline{\mbox{MS}}}_s$, that is evaluated with the SM
renormalization group equations at two-loop order, to the $\DRb$
scheme. The light quark masses, which have only a small
influence on the loop results, have been set to  
\beq
m_u=2.5\,\mev\; , \quad m_d=4.95\,\mev\; , \quad m_s=100\, \mev \quad
\mbox{and} \quad m_c=1.42\, \gev \;. \label{eq:param2}
\eeq
\noindent 
The remaining parameters differ in the two scenarios. Thus we have: \\
\noindent
\underline{Scenario 1:} 
The soft SUSY breaking masses and trilinear couplings are chosen as 
\beq
&&  m_{\tilde{u}_R,\tilde{c}_R} = 
m_{\tilde{d}_R,\tilde{s}_R} =
m_{\tilde{Q}_{1,2}}= m_{\tilde L_{1,2}} =m_{\tilde e_R,\tilde{\mu}_R} = 3\;\mbox{TeV}\, , \;  
m_{\tilde{t}_R}=1909\,\gev \,,\; \non \\ \non
&&  m_{\tilde{Q}_3}=2764\,\gev\,,\; m_{\tilde{b}_R}=1108\,\gev\,,\; 
m_{\tilde{L}_3}=472\,\gev\,,\; m_{\tilde{\tau}_R}=1855\,\gev\,,
 \\ 
&& |A_{u,c,t}| = 1283\,\gev\, ,\; |A_{d,s,b}|=1020\,\gev\,,\; |A_{e,\mu,\tau}| = 751\,\gev\,,\; \\ \non
&& |M_1| = 908\,\gev,\; |M_2|= 237\,\gev\,,\; |M_3|=1966\,\gev \,,\\ \non
&&  \varphi_{A_{d,s,b}}=\varphi_{A_{e,\mu,\tau}}=\varphi_{A_{u,c,t}}=\pi\,,\; 
\varphi_{M_1}=\varphi_{M_2}=\varphi_{M_3}=0
 \;. \label{eq:param4}
\eeq
The remaining input parameters are given by 
\beq
&& |\lambda| = 0.374 \;, \quad |\kappa| = 0.162 \; , \quad |A_\kappa| = 178\,\gev\;,\quad 
|\mu_{\text{eff}}| = 184\,\gev \;, \non \\ 
&&\varphi_{\lambda}= \varphi_\kappa=\varphi_{\mu_{\text{eff}}}=\varphi_u=0\;, \quad \varphi_{A_\kappa}=\pi \;, 
\quad \tan\beta = 7.52 \;,\quad M_{H^\pm} = 1491\,\gev \;.
\eeq
\underline{Scenario 2:}
For the soft SUSY breaking masses and trilinear couplings we chose 
\beq
&&  m_{\tilde{u}_R,\tilde{c}_R} = 
m_{\tilde{d}_R,\tilde{s}_R} =
m_{\tilde{Q}_{1,2}}= m_{\tilde L_{1,2}} =m_{\tilde e_R,\tilde{\mu}_R} = 3\;\mbox{TeV}\, , \;  
m_{\tilde{t}_R}=1170\,\gev \,,\; \non \\ \non
&&  m_{\tilde{Q}_3}=1336\,\gev\,,\; m_{\tilde{b}_R}=1029\,\gev\,,\; 
m_{\tilde{L}_3}=2465\,\gev\,,\; m_{\tilde{\tau}_R}=301\,\gev
 \\ 
&& |A_{u,c,t}| = 1824\,\gev\, ,\; |A_{d,s,b}|=1539\,\gev\,,\; |A_{e,\mu,\tau}| = 1503\,\gev\,,\; \\ \non
&& |M_1| = 862.4\,\gev,\; |M_2|= 201.5\,\gev\,,\; |M_3|=2285\,\gev\\ \non
&&  \varphi_{A_{d,s,b}}=\varphi_{A_{e,\mu,\tau}}=\pi\,,\; 
\varphi_{A_{u,c,t}}=\varphi_{M_1}=\varphi_{M_2}=\varphi_{M_3}=0
 \;. \label{eq:param4scen2}
\eeq
And the remaining input parameters are set as follows, 
\beq
&& |\lambda| = 0.629 \;, \quad |\kappa| = 0.208 \; , \quad |A_\kappa| = 179.7\,\gev\;,\quad 
|\mu_{\text{eff}}| = 173.7\,\gev \;, \non \\ 
&&\varphi_{\lambda}=\varphi_{\mu_{\text{eff}}}=\varphi_u=\varphi_{A_\kappa}=0\;,
\quad \varphi_{\kappa}=\pi \;,  
\quad \tan\beta = 4.02 \;,\quad M_{H^\pm} = 788 \,\gev \;.
\eeq
We follow the SLHA format, which requires $\mu_{\text{eff}}$ as input
parameter. The values for $v_s$ and $\varphi_s$ can then be obtained
by using Eq.~(\ref{eq:mueff}). In the SLHA format, the parameters $\lambda, \kappa,
A_\kappa, \mu_{\text{eff}}, \tan\beta$ as well as the soft 
SUSY breaking masses and trilinear couplings are understood as $\DRb$
parameters at the scale $\mu_R = M_s$\footnote{For $\tan\beta$ this is
only true, if it is read in from the block EXTPAR as done in {\tt
  NMSSMCALC}. Otherwise it is the 
$\DRb$ parameter at the scale $M_Z$.}, whereas the charged Higgs mass is
an OS parameter. We set the SUSY scale $M_s$ to 
\beq
M_s = \sqrt{m_{\tilde Q_3}m_{\tilde t_R}} \;.
\eeq
The resulting supersymmetric particle spectrum from the thus chosen
parameter values is in accordance with present LHC searches for SUSY particles 
\cite{Aad:2012tx,Aad:2012yr,Aad:2013ija,Aad:2014qaa,Aad:2014wea,Aad:2014bva,Aad:2014kra,Chatrchyan:2013xna,CMSPAS13008,Chatrchyan:2013fea,Chatrchyan:2013mya,CMSPAS13018,CMSPAS13019,Khachatryan:2014doa,CMSPAS14011}. Note,
that in the following we will drop the subscript '$\text{eff}$' for
$\mu$. Furthermore, whenever we will use the expressions OS 
and $\DRb$ these refer to the renormalization in the top/stop
sector. 

\subsection{Results  for the loop-corrected self-couplings \label{sec:results1}}
\begin{table}[t]
\begin{center}
 \begin{tabular}{|l||c|c|c|c|c|}
\hline
OS &${H_1}$&${H_2}$&${H_3}$&${H_4}$&${H_5}$\\ \hline \hline
mass tree [GeV] &71.14& 117.49& 211.12& 1491& 1492 \\
main component&$h_u$&$h_s$&$a_s$&$a$&$h_d$\\ \hline
mass one-loop [GeV] &98.65 & 139.17& 217.27& 1490& 1491 \\
main component&$h_s$&$h_u$&$a_s$&$a$&$h_d$\\ \hline
mass two-loop [GeV] &94.68 &125.06 &217.32 &1490 &1491 \\
main component&$h_s$&$h_u$&$a_s$&$a$&$h_d$\\ \hline \hline
%
$\DRb$ &${H_1}$&${H_2}$&${H_3}$&${H_4}$&${H_5}$\\ \hline \hline
mass tree [GeV] &71.14& 117.49& 211.12& 1491& 1492 \\
main component&$h_u$&$h_s$&$a_s$&$a$&$h_d$\\ \hline
mass one-loop [GeV] &91.60 & 120.00 & 217.36 & 1491& 1491 \\
main component&$h_s$&$h_u$&$a_s$&$a$&$h_d$\\ \hline
mass two-loop [GeV] &94.41 &124.24 &217.33 &1490 &1491 \\
main component&$h_s$&$h_u$&$a_s$&$a$&$h_d$\\ \hline 
\end{tabular}
\caption{{\it Scenario 1:} Masses and main components of the neutral
  Higgs bosons at 
  tree and one-loop level and at order ${\cal O}(\alpha_t
  \alpha_s)$ as obtained by using OS (upper) and
  $\DRb$ (lower) renormalization in the top/stop sector.}
\label{tab:massvalues}
\end{center}
\end{table}
In this and the following subsection we discuss the impact of the order
${\cal O}(\al_t \al_s)$ corrections on the trilinear Higgs
self-couplings and on Higgs-to-Higgs decay widths. 
We start by discussing the results for the parameter set called
scenario 1 in the previous subsection.  
The masses of the Higgs bosons and their main composition in terms of
singlet/doublet and scalar/pseudoscalar components at tree level,
one-loop and two-loop order are summarized in
Table~\ref{tab:massvalues} both for the OS and the $\DRb$
renormalization in the top/stop sector. The tree-level stop masses in
this scenario are rather heavy and given by 
\beq
\begin{array}{llll}
\mbox{OS} &:& m_{\tilde{t}_1} = 1992 \mbox{ GeV} \;, \qquad &
m_{\tilde{t}_2} = 2820 \mbox{ GeV} \;, \\
\DRb &:& m_{\tilde{t}_1} = 1911 \mbox{ GeV} \;, \qquad &
m_{\tilde{t}_2} = 2768 \mbox{ GeV} \;.
\end{array}
\eeq
and for the $\DRb$ top mass we have $m_t^{\DRb} = 136.34$~GeV. 
For definiteness, with respect to the mass corrections one-loop means
here and in the following that we include the full EW corrections at
non-vanishing external momenta, while at two-loop level the order
${\cal O}(\alpha_t \alpha_s)$ corrections are computed at vanishing
external momenta. As can be inferred from the table, the
masses of the three lightest scalars are substantially different, so that
mixing effects due to CP-violation for non-vanishing phases cannot be
expected to be significant. The reason for choosing this scenario are
higher order corrections to the trilinear Higgs self-coupling of the
SM-like Higgs boson which are rather important for this parameter
point. This boson is given by the state 
with the largest $h_u$ component and a mass value around 
125~GeV.\footnote{A rather large $h_u$ component is required in order
to reproduce the experimentally measured production rates. They are
mainly due to gluon fusion, which is dominantly mediated by top loops
for small values of $\tan\beta$.} At tree level it is the lightest Higgs
boson $H_1$ that is mainly $h_u$-like, and its mass thus receives large
corrections which are dominantly stemming from the top/stop
sector. The large corrections shift the $H_1$ mass above the one of
$H_2$ so that the two Higgs bosons interchange their roles, as they
are ordered by ascending mass. At one- and two-loop level it is
therefore the second lightest Higgs boson, which is $h_u$-like. 
For convenience, we denote in the following the mass eigenstate that
is dominantly $h_u$-like, by $h$. Furthermore, when we perform
comparisons in the interaction basis at different loop-levels, these
will be done for the $h_u$ state. \s  

\begin{figure}[t]
\hspace*{-0.4cm}
 \includegraphics[width=0.53\textwidth]{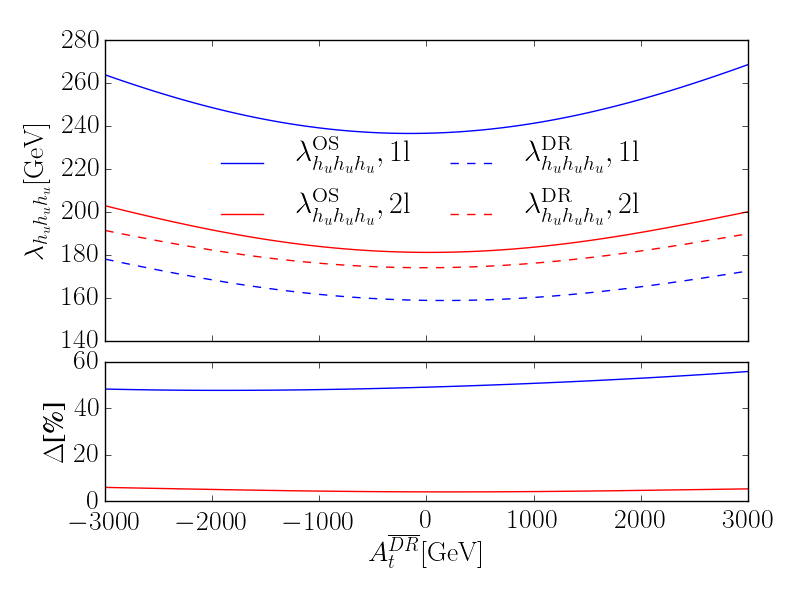} 
\hspace*{-0.6cm} 
\includegraphics[width=0.53\textwidth]{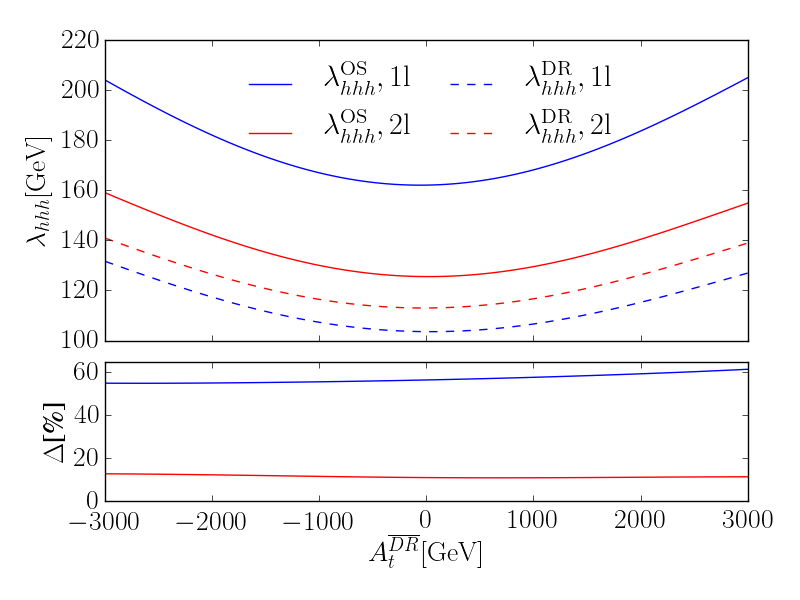}
\caption{{\it Scenario 1:} Upper panels: Trilinear self-coupling of the
  $h_u$ interaction state (left) and the mass eigenstate $h$ (right)
  as a function of $A_t^{\overline{\text{DR}}}$ including the one-loop
  correction (blue/two outer lines) and also the two-loop corrections
  (red/two middle lines). In the top/stop sector either the OS scheme
  (solid lines) or the $\overline{\mbox{DR}}$ scheme (dashed lines)
  has been applied. Lower Panels: Absolute value of the relative
  deviation of the correction using OS or ${\overline{\mbox{DR}}}$
  renormalization in the top/stop sector, $\Delta =
  |\lambda_{HHH}^{m_t(\overline{\text{DR}})}-\lambda_{HHH}^{m_t
    \text{(OS)}}|/\lambda_{HHH}^{m_t(\overline{\text{DR}})}$ 
  ($H= h_u, h$), in percent as a function of
  $A_t^{\overline{\text{DR}}}$, at one-loop (blue/upper) and two-loop
  (red/lower).} \label{fig:atvariatonscen1}
\end{figure}
In Fig.~\ref{fig:atvariatonscen1} we show the dependence of the
one- and two-loop corrections to the Higgs self-coupling 
on the $\DRb$ parameter $A_t$ in the two different renormalization
schemes applied in the 
stop sector. The one-loop corrections have been obtained at order
${\cal O}(\alpha_t)$ for vanishing external momenta. We explicitly
verified, that the differences between the one-loop result in this
approximation and the one including the full one-loop corrections for
non-vanishing momenta at the threshold\footnote{The
  non-vanishing momenta 
  at the threshold have been set to $p_2^2=p_3^2=m_h^2$ for two of the
  external momenta and to $p_1^2=4m_h^2$ for the remaining one. Here
  $m_h$ denotes the two-loop corrected mass value of the SM-like Higgs
boson.} are below 4\% for the 
investigated parameter points. Two-loop corrections always refer to the order
${\cal O}(\alpha_t \alpha_s)$ corrections at vanishing external
momenta. The left plot of Fig.~\ref{fig:atvariatonscen1} shows the
corrections to the self-coupling of $h_u$ in the
interaction basis, $\lambda_{h_u h_u
  h_u}$. Figure~\ref{fig:atvariatonscen1} (right) displays the   
loop-corrected self-couplings after rotation to the mass eigenstate
$h$ with dominant $h_u$ component. The rotation to the mass
eigenstates is performed with the mixing matrix ${\cal R}^{(2)}$
defined in Eq.~(\ref{eq:phirotfull}) for both the one- and the two-loop curves
in the plot. The mass values and mixing matrix
elements have been computed with {\tt NMSSMCALC}. Note that at
two-loop order the $h_u$ dominated state is given by the second
lightest Higgs boson $H_2$, {\it cf.}~Table~\ref{tab:massvalues}. The
dependence on $A_t$ is more pronounced after 
rotation to the mass eigenstates. Overall, however, the size and shape
of the corrections both in the interaction and in the mass eigenstates
are comparable. At the parameter point of scenario 1 the
tree-level coupling $\lambda_{h_u h_u h_u} = 
101.70$~GeV in both renormalization schemes. In the OS scheme the
one-loop correction increases it by 140\% while it is decreased by
24\% to two-loop order. In the $\DRb$ scheme the increase is of 74\%
going from tree- to one-loop order supplemented by another increase of
9\% when adding the two-loop corrections. The reason,
why the one- and 
two-loop corrections differ much more in the OS scheme than in the
$\overline{\mbox{DR}}$ scheme can be understood as follows. In
the $\overline{\mbox{DR}}$ scheme the top quark mass, which according to
the SLHA accord is an OS parameter, has to be converted to the
$\overline{\mbox{DR}}$ value. Thereby, the finite counterterm to the
top mass, which in the OS scheme is included at two-loop level, is
already induced at one-loop level in the value of the
$\overline{\mbox{DR}}$ mass. In this way some corrections of order
${\cal O}(\alpha_t \alpha_s)$, which in the OS scheme only appear at
the two-loop level, are moved to the one-loop level, {\it cf.}~also
\cite{Muhlleitner:2014vsa}. \s

The lower panels of Fig.~\ref{fig:atvariatonscen1} display the difference in the
self-couplings when using the two different renormalization schemes in the
top/stop sector,
\beq
\Delta =
\frac{|\lambda_{HHH}^{m_t(\overline{\text{DR}})}-\lambda_{HHH}^{m_t
    \text{(OS)}}|}{\lambda_{HHH}^{m_t(\overline{\text{DR}})}} \;,
\eeq
where $H$ both refers to the $h_u$ dominated mass eigenstate $h$, and to
the $h_u$ interaction eigenstate. This value gives a rough estimate of
the theoretical error in the Higgs self-coupling due to the unknown higher
order corrections. In the interaction eigenstate it is of order
${\cal O}(50\%)$ at one-loop level, decreasing to roughly 4\% at
two-loop level. In the mass eigenstate it is about 5\% higher at
both loop orders. The inclusion of the two-loop corrections hence
substantially decreases the theoretical uncertainty.  \s

\begin{figure}[t]
\hspace*{-0.4cm}
 \includegraphics[width=0.53\textwidth]{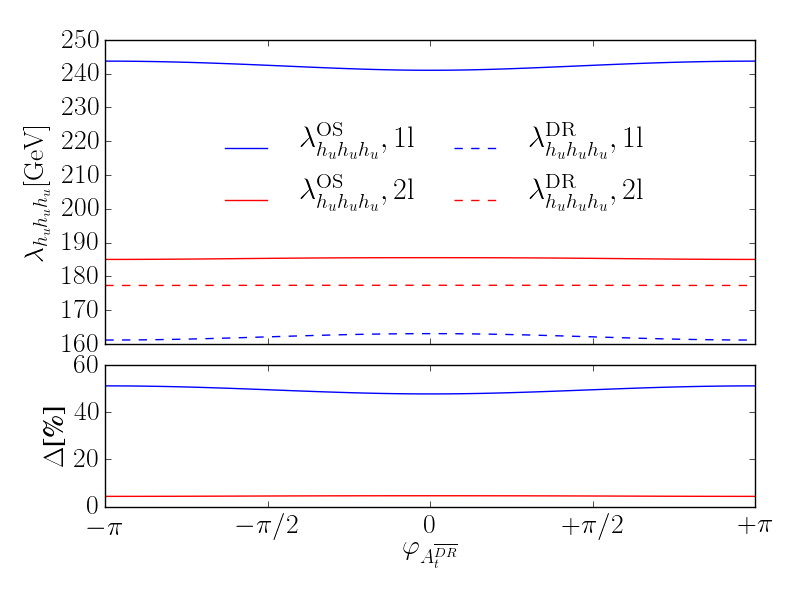} 
\hspace*{-0.6cm}
\includegraphics[width=0.53\textwidth]{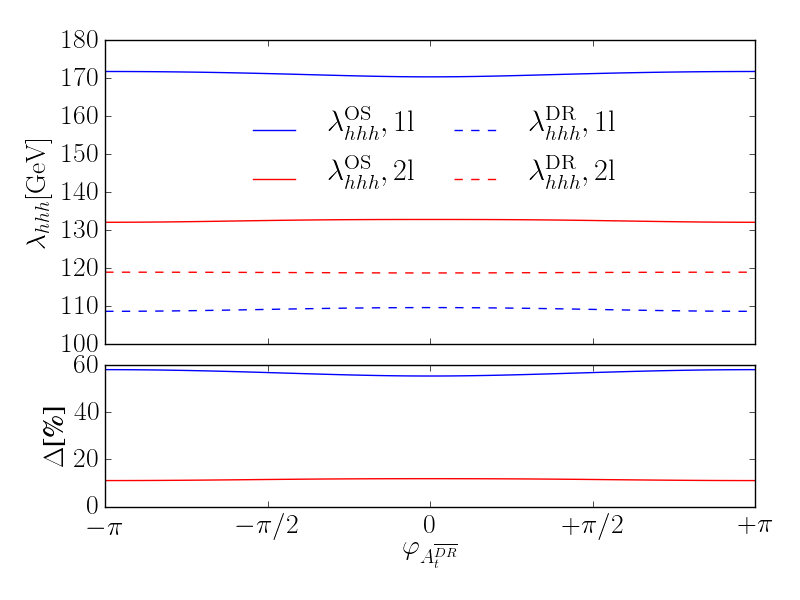}
\caption{{\it Scenario 1:} Same as Fig.~\ref{fig:atvariatonscen1}, but
  now as a function of
  $\varphi_{A_t}^{\overline{\text{DR}}}$. \label{fig:phiatvariationscen1}}  
\end{figure}
Figure~\ref{fig:phiatvariationscen1} shows the same as
Fig.~\ref{fig:atvariatonscen1} but now as a function of the phase
$\varphi_{A_t}$. All other CP-violating phases have been kept to
zero. The figure shows that the dependence of the loop corrections on
the phase is almost negligible, as expected for radiatively induced
CP-violation. The size of the loop corrections and the remaining
theoretical uncertainty are of the same order as for the variation of
$A_t$. \s

\begin{table}[t]
\begin{center}
 \begin{tabular}{|l||c|c|c|c|c|}
\hline
OS &${H_1}$&${H_2}$&${H_3}$&${H_4}$&${H_5}$\\ \hline \hline
mass tree [GeV] &79.15& 103.55& 146.78& 796.62& 803.86 \\
main component&$h_s$&$h_u$&$a_s$&$h_d$&$a$\\ \hline
mass one-loop [GeV] &103.45 & 129.15& 139.83& 796.53& 802.94 \\
main component&$h_s$&$a_s$&$h_u$&$h_d$&$a$\\ \hline
mass two-loop [GeV] &102.99 &126.09 &128.94 &796.45 &803.07 \\
main component&$h_s$&$h_u$&$a_s$&$h_d$&$a$\\ \hline \hline
%
$\DRb$ &${H_1}$&${H_2}$&${H_3}$&${H_4}$&${H_5}$\\ \hline \hline
mass tree [GeV] &79.15& 103.55& 146.78& 796.62& 803.86 \\
main component&$h_s$&$h_u$&$a_s$&$h_d$&$a$\\ \hline
mass one-loop [GeV] &102.80 & 120.52 & 128.80 & 796.36& 803.09 \\
main component&$h_s$&$h_u$&$a_s$&$h_d$&$a$\\ \hline
mass two-loop [GeV] &103.09 &124.55 &128.91 &796.36 &803.03 \\
main component&$h_s$&$h_u$&$a_s$&$h_d$&$a$\\ \hline 
\end{tabular}
\caption{{\it Scenario 2:} Masses and main components of the neutral
  Higgs bosons at 
  tree and one-loop level and at order ${\cal O}(\alpha_t
  \alpha_s)$ as obtained by using OS (upper) and
  $\DRb$ (lower) renormalization in the top/stop sector.}
\label{tab:massvalues2}
\end{center}
\end{table} 
We now turn to the discussion of scenario 2. The masses and dominant
composition of the mass eigenstates at tree level, one- and two-loop order are
summarized in Table \ref{tab:massvalues2}. In the OS scheme again the
composition of the mass ordered states changes when going from
tree level to one-loop level and from one- to two-loop level.
In contrast to scenario 1 the masses of $H_2$ and $H_3$ 
are now much closer together, in particular after inclusion of the
two-loop corrections. We therefore expect CP-violating effects to be
more important here. The $H_2$ state is identified with the
discovered Higgs boson. The stop masses are again rather heavy with
\beq
\begin{array}{llll}
\mbox{OS} &:& m_{\tilde{t}_1} = 1145 \mbox{ GeV} \;, \qquad &
m_{\tilde{t}_2} = 1421 \mbox{ GeV} \;, \\
\DRb &:& m_{\tilde{t}_1} = 1126 \mbox{ GeV} \;, \qquad &
m_{\tilde{t}_2} = 1387 \mbox{ GeV} \;.
\end{array}
\eeq

\begin{figure}[t!]
\hspace*{-0.1cm}
\hspace*{-0.6cm}
 \includegraphics[width=0.53\textwidth]{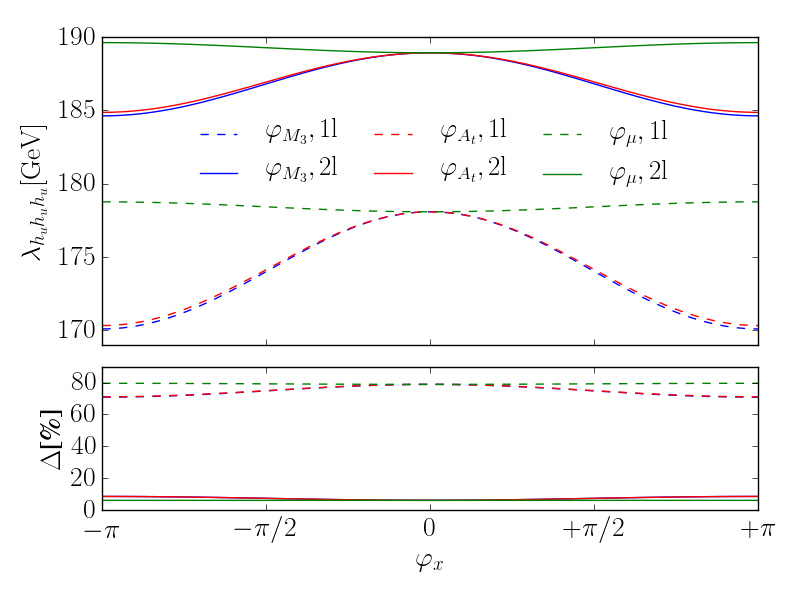} 
\hspace*{-0.4cm}
\includegraphics[width=0.53\textwidth]{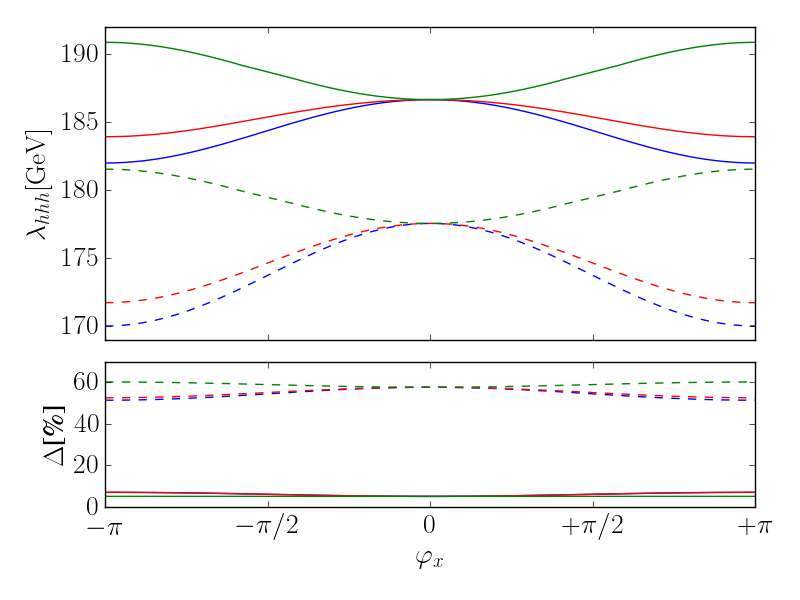}
\caption{{\it Scenario 2:} Upper panels: Trilinear self-coupling of the
  $h_u$ interaction state (left) and the mass eigenstate $h$ (right)
  at one-loop (dashed lines) and two-loop (solid lines) as a function of the phases
  $\varphi_{\mu}$ (green/grey), $\varphi_{A_t}$ (red/black upper) and $\varphi_{M_3}$
  (blue/black lower).  
   Lower Panels: Size of the relative correction of $n^{th}$
  order to the Higgs self-coupling with respect to the
  $(n-1)^{th}$ order -- {\it
    i.e.}~$\Delta=|\lambda_{HHH}^{(n)} -
  \lambda_{HHH}^{(n-1)}|/\lambda_{HHH}^{(n-1)} $ -- in percent for
  $H=h_u$ (left) and $H=h$ (right) as a function of the phases
  $\varphi_{\mu}$ (green/grey), $\varphi_{A_t}$ (red/black) and
  $\varphi_{M_3}$ (blue/black) for $n=2$ (solid 
  line) and $n=1$ (dashed line). The red and blue lines almost lie on
  top of each other. In the top/stop sector we have
  applied $\DRb$ renormalization.}
\label{fig:lambdaphasevariation}
\end{figure}
In Fig.~\ref{fig:lambdaphasevariation} we show the dependence of the
Higgs self-coupling of the $h_u$ state in the interaction basis
(left) and of the $h_u$-like mass eigenstate $h$ (right) 
at one- (dashed) and two-loop order (full) for $\DRb$ renormalization
in the top/stop sector as a function of the phases $\varphi_{M_3}$,
$\varphi_{A_t}$ and $\varphi_\mu$. For illustrative purposes we have
varied the phases in  
rather large ranges although they might already be excluded by
experiment. We start from our 
original CP-conserving scenario and turn on the phases one by one. Note, that
$\varphi_\mu$ has been varied such that the CP-violating phase 
$\varphi_y= \varphi_\kappa - \varphi_\lambda + 2\varphi_s -
\varphi_u$, that appears already at tree level in the Higgs sector,
remains zero, {\it i.e.}~$\varphi_\lambda$ and $\varphi_s$ were varied
at the same time as $\varphi_\lambda= 2 \varphi_s = 2/3 \varphi_\mu$,
while $\varphi_\kappa$ and $\varphi_u$ are kept zero. 
As expected, the loop-corrected couplings show a somewhat larger
dependence on $\varphi_{A_t}$ than in scenario 1, in particular in the
mass eigenstate basis. Defining as
\beq
\delta \lambda_{HHH} = \frac{\lambda_{HHH} (\pi) - \lambda_{HHH}
  (0)}{\lambda_{HHH} (0)} \;,
\eeq
we have in the mass eigenstate basis $H \equiv h$ the variations 
\beq
\delta \lambda_{hhh}^{\varphi_\mu} = 2.2\% \;, \quad
\delta \lambda_{hhh}^{\varphi_{A_t}} = 1.6\% \quad \mbox{and} \quad
\delta \lambda_{hhh}^{\varphi_{M_3}} = 2.7\% 
\eeq
for the two-loop corrected self-coupling. Note, that the one-loop
corrected self-couplings show a dependence on the phase of $M_3$,
although the genuine diagrammatic gluino corrections only appear at
two-loop level. This dependence enters through the conversion of the
OS top quark mass to the $\overline{\mbox{DR}}$ mass. Overall, the
dependence of the loop corrected self-couplings on the CP-violating
phases is smaller in the interaction states than in the mass
eigenstates, which are obtained by rotating the interaction
states with the mixing elements obtained from the loop corrected
masses, that also depend on the CP-violating phases. \s

The lower panels show the relative corrections, defined at order $n=1,2$ as 
\beq
\Delta = \frac{|\lambda_{HHH}^{(n)} -
  \lambda_{HHH}^{(n-1)}|}{\lambda_{HHH}^{(n-1)}}  \;.
\eeq
In the interaction basis they are of order $\sim 70-80$\% for the
one-loop corrections relative to the tree-level coupling and are
somewhat larger than the corresponding values in the mass eigenstate
basis, which are of order $\sim 50-60$\%. For the two-loop coupling 
relative to the one-loop coupling the corrections are
significantly reduced to about $5-8$\% in both the interaction and the
mass eigenstate basis. The two-loop corrections hence considerably
reduce the theoretical uncertainty. \s

\begin{figure}[t!]
\hspace*{-0.6cm}
\includegraphics[width=0.53\textwidth]{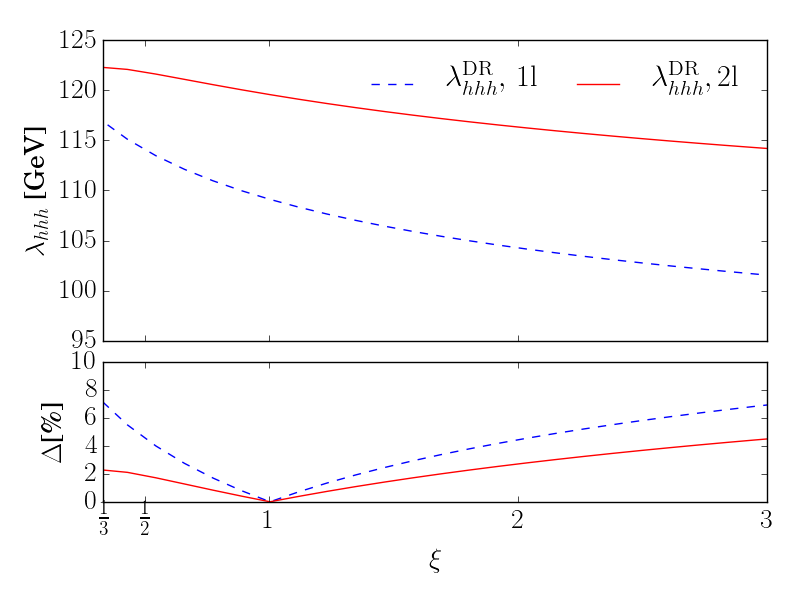} 
\hspace*{-0.4cm}
\includegraphics[width=0.53\textwidth]{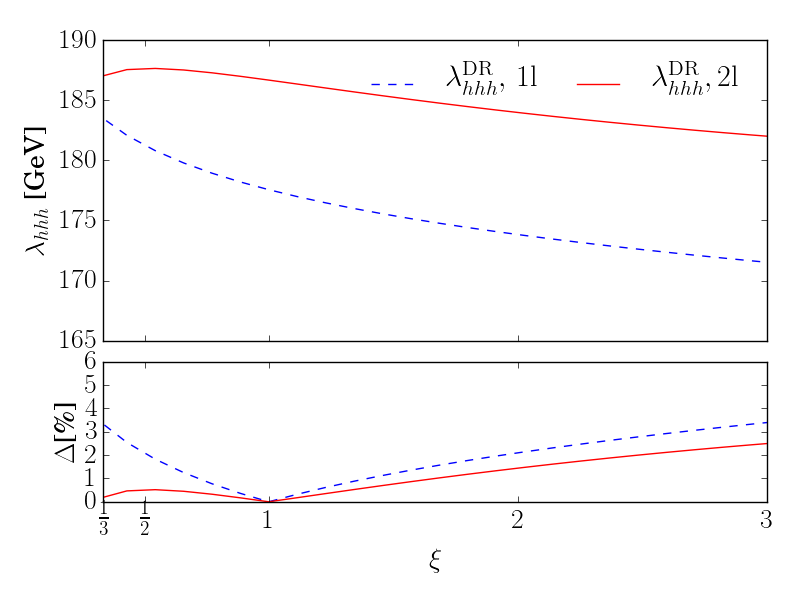}
\caption{Scale dependence of the trilinear coupling
  $\lambda_{hhh}$ at one- (blue/dashed) and two-loop order
  (red/full) for {\it scenario 1} (left) and {\it scenario 2}
  (right). The scale $\mu_R=\xi M_s$ has been varied in an interval of
  $\xi=1/3...3$ around the central scale $\mu_0 = M_s$. The lower plots show
  the variation in percent compared to the central scale, {\it i.e.}~$\Delta =
  [\lambda_{hhh} (\mu_R) - \lambda_{hhh} (\mu_0)]/\lambda_{hhh}
  (\mu_0)$.}
\label{fig:scalevariation}
\end{figure}
In order to further study the theoretical uncertainty, we show in
Fig.~\ref{fig:scalevariation} for scenario 1 (left) and scenario 2
(right) the scale variation of the trilinear Higgs self-coupling in the
mass eigenstate $h$ at one- and at two-loop order. We have
varied the renormalization scale $\mu_R$ between 1/3 and 3 times the central
scale $\mu_0 = M_s$. The scale variation affects the
$\overline{\mbox{DR}}$ parameters entering the calculation. In the
absence of an implementation of the 2-loop renormalization group
equations (RGE) for the complex NMSSM, which is devoted to future work, we
obtain the parameters at the different scales by exploiting the
relation between $\overline{\mbox{DR}}$ and OS parameters, as
explained in Appendix~\ref{app:trick}. This should approximate the
results obtained from the RGE running rather well, in case the scale is
not varied in a too large range. Since the scale variation provides only
a rough estimate of the error made by neglecting higher order
corrections this approach is sufficient for our purpose. As can be
inferred from the figures, in scenario 1 the one-loop coupling is altered by up
to 7\% compared to its value at the central scale in the investigated
range. This reduces to 2-5\% at two-loop order. In scenario 2 the
corresponding numbers at one-loop order are 3.5\% compared to up to
2.5\% for the two-loop coupling. As expected, the scale dependence
reduces when going from one- to two-loop order. Note, however, that
these numbers should not be taken as estimate for the residual
theoretical uncertainty.

\subsection{Phenomenological implications \label{sec:results2}}
We now turn to the discussion of the phenomenological implications due
to the loop-corrected Higgs self-couplings. Higgs self-couplings are involved in
Higgs-to-Higgs decays and in Higgs pair production processes. At the
LHC, pair production dominantly proceeds through gluon fusion. This
process, however, includes EW corrections  
beyond those approximated by the loop-corrected effective trilinear
couplings. As they are not available at present we will not discuss Higgs pair
production further and concentrate on Higgs-to-Higgs decays. \s

The decay width for the Higgs-to-Higgs decay $H_i \to H_j H_k$
including the two-loop corrections to the Higgs self-coupling is
obtained from  
\beq
\Gamma (H_i \to H_j H_k) = \frac{\lambda^{1/2} (M_{H_i}^2,M_{H_j}^2,
  M_{H_k}^2)}{16 \pi f \, M_{H_i}^3} |{\cal M}_{H_i \to H_j H_k}|^2
\;, 
\label{eq:partialwidth}
\eeq
where $f=2$ for identical final state particles and $f=1$
otherwise. The decay amplitude is denoted by ${\cal M}_{H_i \to H_j
  H_k}$ and $\lambda=(x-y-z)^2-4yz$ is the two-body phase space
function. In case of CP-violation all Higgs-to-Higgs decays between
the five neutral Higgs bosons are possible, if kinematically allowed, so that
$i,j,k=1,...,5$. In the CP-conserving case, however, only the trilinear Higgs
couplings between three CP-even or one CP-even and two CP-odd Higgs
bosons are non-vanishing. The matrix element is given by 
\beq
{\cal M}_{H_i \to H_j H_k} 
&=& \sum_{i',j',k'=1}^5 {\bf Z}_{ii'} {\bf
  Z}_{jj'} {\bf Z}_{kk'} \Gamma_{h_{i'} h_{j'} h_{k'}} + \delta M_{H_i \to H_j
  H_k}^{\text{mix}} \;. \label{eq:mateldecay}
\eeq
Here $\Gamma_{h_{i'} h_{j'} h_{k'}}$ is the loop
corrected trilinear Higgs self-coupling in the tree-level mass
eigenstate basis, where the Goldstone boson has been singled out. We
here include at one-loop level the full electroweak corrections
\cite{Nhung:2013lpa} at $p^2 \ne 0$, where we set the 4-momenta of the
external Higgs particles equal to the respective loop-corrected Higgs
mass values $p_{H_{i,j,k}}^2 = M_{H_{i,j,k}}^2$ as obtained with {\tt NMSSMCALC}
\cite{Ender:2011qh,Graf:2012hh,Muhlleitner:2014vsa,Baglio:2013iia},
and at two-loop the order ${\cal O}(\alpha_t \alpha_s)$
corrections at $p^2 =0$.\footnote{In the loops the tree-level masses
  for the Higgs bosons are used to ensure the cancellation of the UV
  divergences.} The proper on-shell conditions of the external Higgs
bosons as required in the decay process are ensured by rotating the
tree-level mass eigenstates $h_{i',j',k'}$ to the loop
corrected mass eigenstates $H_{i,j,k}$ with the matrix ${\bf Z}$, {\it
  cf.}~\cite{Ender:2011qh,Nhung:2013lpa}. In this calculation we include at
one-loop order 
the full electroweak corrections at non-vanishing external momenta. At
two-loop order as usual the order ${\cal O} (\alpha_t \alpha_s)$
corrections which are available only at $p^2 =0$ are taken into
account.\footnote{As we 
  investigate here the decay 
  of heavy particles in the initial and final states, it makes sense
  not to work in the zero momentum approximation if possible and
  include at one-loop level the full momentum dependence.} The $\delta
M_{H_i \to H_j H_k}^{\text{mix}}$ accounts for the 
contributions stemming from the mixing of the CP-odd components of the
external Higgs bosons with the Goldstone and with
the $Z$ boson, respectively. These contributions, which are evaluated
by setting the external momenta to the tree-level masses in order to
maintain gauge invariance, are small already at one-loop order 
compared to the remaining contributions to the decay amplitude, as has
been shown in \cite{Nhung:2013lpa}. We hence do not include the
two-loop contributions, that can safely be expected to be
negligible. \s

In the plots below we show apart from the two-loop corrected decay
widths also the ones at one-loop order. The only change required to
adapt formula (\ref{eq:mateldecay}) to this case is in $\Gamma_{h_{i'}
  h_{j'} h_{k'}}$ where solely the one-loop corrections to the vertex 
functions together with the corresponding counterterms are included. In
particular we also use the two-loop corrected mass eigenstates and
mixing matrix elements for the external particles (apart from the
mixing contribution with the Goldstone and $Z$ boson of course). \s

The scenario which the following discussion is based on and which
has been checked to be compatible with the constraints from the LHC Higgs
data, is given by: \s

\noindent
\underline{Scenario 3:} 
The soft SUSY breaking masses and trilinear couplings are chosen as 
\beq
&&  m_{\tilde{u}_R,\tilde{c}_R} = 
m_{\tilde{d}_R,\tilde{s}_R} =
m_{\tilde{Q}_{1,2}}= m_{\tilde L_{1,2}} =m_{\tilde e_R,\tilde{\mu}_R} = 3\;\mbox{TeV}\, , \;  
m_{\tilde{t}_R}=1940\,\gev \,,\; \non \\ \non
&&  m_{\tilde{Q}_3}=2480\,\gev\,,\; m_{\tilde{b}_R}=1979\,\gev\,,\; 
m_{\tilde{L}_3}=2667\,\gev\,,\; m_{\tilde{\tau}_R}=1689\,\gev\,,
 \\ 
&& |A_{u,c,t}| = 1192\,\gev\, ,\; |A_{d,s,b}|=685\,\gev\,,\; |A_{e,\mu,\tau}| = 778\,\gev\,,\; \\ \non
&& |M_1| = 517\,\gev,\; |M_2|= 239\,\gev\,,\; |M_3|=1544\,\gev \,,\\ \non
&&  \varphi_{A_{d,s,b}}=\varphi_{A_{e,\mu,\tau}}=0 \,, \;\varphi_{A_{u,c,t}}=\pi\,,\; 
\varphi_{M_1}=\varphi_{M_2}=\varphi_{M_3}=0
 \;. \label{eq:param4scen3}
\eeq
The remaining input parameters are given by 
\beq
&& |\lambda| = 0.267 \;, \quad |\kappa| = 0.539 \; , \quad |A_\kappa| = 810\,\gev\;,\quad 
|\mu_{\text{eff}}| = 104\,\gev \;, \non \\ 
&&\varphi_{\lambda}= \varphi_\kappa=\varphi_{\mu_{\text{eff}}}=\varphi_u=0\;, \quad \varphi_{A_\kappa}=\pi \;, 
\quad \tan\beta = 8.97 \;,\quad M_{H^\pm} = 613\,\gev \;.
\eeq
This results in the Higgs mass spectrum given in
Table~\ref{tab:decayspectr} at tree, one- and two-loop level together
with the main singlet/doublet and scalar/pseudoscalar components at
each loop level.
\begin{table}[t!]
\begin{center}
 \begin{tabular}{|l||c|c|c|c|c|}
\hline
OS &${H_1}$&${H_2}$&${H_3}$&${H_4}$&${H_5}$\\ \hline \hline
mass tree [GeV] &49.17& 99.83& 609.21& 611.77& 715.92 \\
main component&$h_s$&$h_u$&$a$&$h_d$&$a_s$\\ \hline
mass one-loop [GeV] &87.36 & 139.10& 608.71& 611.37& 694.73 \\
main component&$h_s$&$h_u$&$a$&$h_d$&$a_s$\\ \hline
mass two-loop [GeV] &83.66 &124.95 &608.73 &611.37 &694.76 \\
main component&$h_s$&$h_u$&$a$&$h_d$&$a_s$\\ \hline \hline
%
$\DRb$ &${H_1}$&${H_2}$&${H_3}$&${H_4}$&${H_5}$\\ \hline \hline
mass tree [GeV] &49.17& 99.83& 609.21& 611.77& 715.92 \\
main component&$h_s$&$h_u$&$a$&$h_d$&$a_s$\\ \hline
mass one-loop [GeV] &80.66 & 119.68 & 608.72 & 611.37& 694.79 \\
main component&$h_s$&$h_u$&$a$&$h_d$&$a_s$\\ \hline
mass two-loop [GeV] &83.03 &124.34 &608.71 &611.36 &694.78 \\
main component&$h_s$&$h_u$&$a$&$h_d$&$a_s$\\ \hline 
\end{tabular}
\caption{{\it Scenario 3:} Masses and main components of the neutral
  Higgs bosons at 
  tree and one-loop level and at order ${\cal O}(\alpha_t
  \alpha_s)$ as obtained by using OS (upper) and
  $\DRb$ (lower) renormalization in the top/stop sector.}
\label{tab:decayspectr}
\end{center}
\end{table}
\begin{figure}[b!]
\includegraphics[width=0.5\textwidth]{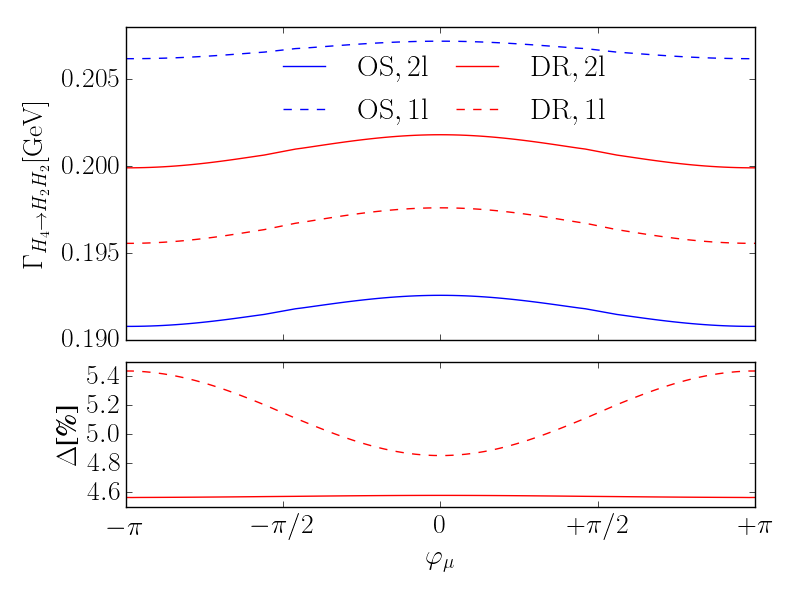} \includegraphics[width=0.5\textwidth]{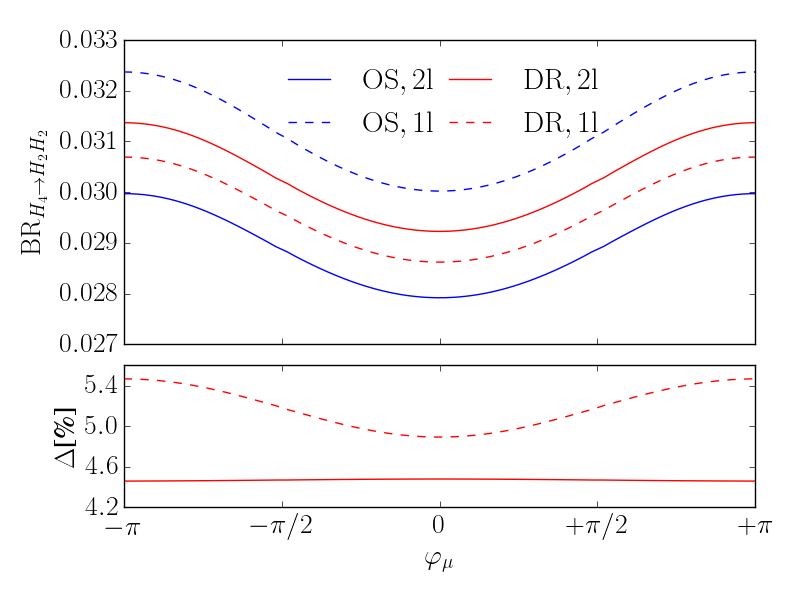}
\caption{{\it Scenario 3:} Upper: Loop-corrected decay width (left) and
  branching ratio (right) of the Higgs-to-Higgs decay $H_4 \to H_2
  H_2$ as a function of $\varphi_\mu$ with the top/stop sector
  renormalized in the OS (blue/two outer lines) and in the
  $\overline{\mbox{DR}}$ scheme (red/two inner lines) at 1-loop
  (dashed) and at 2-loop (full) order. Lower: Relative deviation
  between the two renormalization schemes $\Delta =
  |O(\overline{\mbox{DR}}) - O(\mbox{OS})|/O(\overline{\mbox{DR}})$
  for $O=\Gamma$ (left) and $O=\mbox{BR}$ (right) at 1-loop (dashed) and
  2-loop (full).}
\label{fig:widthbrh4h2h2}
\end{figure}

In Fig.~\ref{fig:widthbrh4h2h2} (left) we show the partial decay width
for the decay of the heavy $h_d$-like 
Higgs boson $H_4$ into a pair of SM-like Higgs bosons $H_2$, including
the higher order corrections to the Higgs-self-couplings at one- and
two-loop level as obtained from Eq.~(\ref{eq:partialwidth}). We start
from the parameter point of scenario 3 with vanishing phase
$\varphi_\mu=0$. The phase is then varied in the range $-\pi...\pi$,
such that at tree level the CP-violating phase $\varphi_y$ in the
Higgs sector vanishes. As expected the
dependence of the decay width on the CP-violating phase induced
through the loop corrections is small, remaining below the per-cent
level. For $\varphi_\mu=0$ the tree-level decay width in the OS scheme
is 0.171~GeV and 0.186~GeV in the $\overline{\mbox{DR}}$
scheme.\footnote{Note, that of course also in the tree-level decay
  width we use the $H_4$ and $H_2$ mass values 
  including the two-loop corrections and rotate to the mass
  eigenstates with the corresponding mixing matrix elements.} In
the latter the one-loop corrections increase the decay width by 6.5\% 
and the two-loop corrections add another 2.0\% on top of that. In the OS we find
a 21\% increase at one-loop and a 7\% decrease at two-loop so that at
two-loop order the results in the two renormalization schemes approach
each other, as can also be inferred from the lower left plot, which shows
that the dependence on the renormalization scheme decreases from one-
to two-loop level. Depending on the renormalization scheme different 
input parameters of the top/stop sector have to be converted to match
the required renormalization scheme. The corresponding induced two-loop
corrections into the one-loop corrections lead to a different
dependence on the CP-violating phase of the two schemes at one-loop
level. At two-loop level this difference in the phase dependence is
then almost washed out and the scheme dependence is about 4.58\%
independent of $\varphi_\mu$. 
For the computation of the branching ratio of the decay, shown in
Fig.~\ref{fig:widthbrh4h2h2} (right), we replace in the
program package {\tt NMSSMCALC} the tree-level decay widths $\Gamma
(H_i\to H_j H_k)$ with our loop-corrected ones. The branching ratio,
which with ${\cal O}(3\%)$ is very small shows the same trend as the
decay width with respect to the loop corrections. \s

The non-vanishing CP-violating phase induces through the higher order
corrections CP mixing in the Higgs mass eigenstates, such that
otherwise not allowed decays of {\it e.g.}~the CP-odd doublet-like
$H_3$ into a pair of SM-like $H_2$ bosons are possible. The branching
ratio remains, however, tiny, reaching at most 0.58 per mille for
$|\varphi_\mu| \approx \pi/2$ in our scenario. 

\section{Conclusions \label{sec:conclusion}}
The search for New Physics and the proper interpretation of the
experimental data requires from the theoretical side precise
predictions of parameters and observables. In this work we 
computed the two-loop corrections to the trilinear Higgs
self-couplings in the CP-violating NMSSM. Originating from the Higgs
potential the Higgs boson masses and self-couplings are related to
each other. For a consistent interpretation therefore the level of
accuracy of the self-couplings has to match the one of the masses,
that have been provided previously up to the two-loop level. Here, 
the two-loop corrections to the self-couplings have been calculated at
the same order ${\cal O}(\alpha_t \alpha_s)$ at vanishing external
momenta. We have allowed for two 
renormalization schemes in the top/stop sector, namely OS and $\DRb$
renormalization. Depending on the scenario and the renormalization
scheme, the two-loop corrections are of the order of 5-10\% relative
to the one-loop couplings, compared to up to 80\% for the one-loop
corrections relative to the tree-level values. The investigation of the remaining
theoretical uncertainty performed by varying the renormalization
scheme of the top/stop sector or by changing the renormalization scale
confirmed that the theoretical error is reduced through the inclusion
of the two-loop corrections.  As expected the dependence on the
CP-violating phase due to radiatively induced CP-violation is small
and of the order of a few percent. \s

The trilinear self-couplings are relevant in Higgs pair production
processes and Higgs-to-Higgs decays, which now become accessible at
run 2 of the LHC. While Higgs pair production requires the inclusion
of further higher order corrections beyond the loop-corrected Higgs
self-couplings provided in this work, the inclusion of the radiatively
corrected trilinear Higgs self-coupling improves the prediction for the
Higgs-to-Higgs decay rates and related branching ratios. For the
investigated scenario and decay we find that the two-loop corrections alter the
decay width by 2\% (7\%) in the $\DRb$ (OS) scheme
with respect to the one-loop level, which is to be compared to about
7\% (21\%) when going from tree- to one-loop level. The dependence on
the renormalization scheme is reduced from $\sim 4.8-5.4$\% 
in the investigated range of the phase $\varphi_\mu$ at one-loop level
to $\sim 4.5$\% at two-loop level. The behaviour in the branching ratio
is similar to the one of the decay width. \s

In summary, the inclusion of the two-loop corrections at ${\cal O}(\alpha_t
\alpha_s)$ in the approximation of vanishing external momenta in the
trilinear Higgs self-couplings of the CP-violating NMSSM Higgs sector
is necessary to match the available precision in the Higgs masses and
to allow for a consistent interpretation of the Higgs data. Being of
the order of 10\% they have been shown to further reduce the theoretical error
due to missing higher order corrections. 

\subsubsection*{Acknowledgments} 
MM and DTN have been supported in part by the DFG SFB/TR9
``Computational Particle Physics''. HZ acknowledges financial support
from the Graduiertenkolleg ``GRK 1694: Elementarteilchenphysik bei
h\"ochster Energie und h\"ochster Pr\"azision''.
The authors thank Michael Spira and Kathrin Walz for helpful discussions.

\section*{Appendix}
\begin{appendix}
\section{Tree-level trilinear Higgs self-couplings \label{sec:tree-leve-trilinear}} 
In this appendix we present the tree-level trilinear Higgs
self-couplings in the interaction eigenstates, $\lambda_{ijk},$ $i,j,k=1,...,6$
with the correspondences  $1\mathrel{\widehat{=}}h_d$,
$2\mathrel{\widehat{=}}h_u$, $3\mathrel{\widehat{=}}h_s$,
$4\mathrel{\widehat{=}}a_d$, $5\mathrel{\widehat{=}}a_u$,
$6\mathrel{\widehat{=}}a_s$. They are symmetric in the
  three indices. Using the short-hand notations $c_x\equiv \cos x$ etc. and 
\bea
\phi_x&=&\varphi_u + \varphi_s+ \varphi_\lambda +
\varphi_{A_\lambda}\,, \nonumber \\
\phi_y&=&\varphi_u -2 \varphi_s+ \varphi_\lambda - \varphi_{\kappa}\,,\\
\phi_z&=&3 \varphi_s + \varphi_{A_\kappa}+ \varphi_\kappa\,, \nonumber
\eea 
we have
\begin{align}
\lambda_{111}&=\fr{3c_\beta M_Z^2}{v}\,, \quad \lambda_{112}=-\fr{s_\beta M_Z^2}{v}+|\lambda|^2 s_\beta v\,, \quad \lambda_{113}= |\lambda|^2  v_s\,,\quad
\lambda_{114}= 0\,, \quad \lambda_{115}=0\,, \nonumber \\
 \lambda_{116}&=0\,,\quad
\lambda_{122}= -\fr{c_\beta M_Z^2}{v}+ |\lambda|^2 c_\beta v\,, \quad\lambda_{123}=-\fr{|A_\lambda| |\lambda|c_{\phi_x}}{\sqrt{2}} - |\lambda||\kappa| v_sc_{\phi_y}\,, \quad \lambda_{124}=0\,,\nonumber \\
\lambda_{125}&=0 \,, \quad \lambda_{126}=-\fr32|\lambda||\kappa| v_s s_{\phi_y}\,, \quad \lambda_{133}= |\lambda|^2 v c_\beta - |\kappa| |\lambda|v  s_\beta c_{\phi_y}\,,\quad \lambda_{134}= 0\,,
 \nonumber\\
 \lambda_{135}&= \fr12 |\lambda||\kappa| v_s s_{\phi_y}\,, \quad \lambda_{136}=-|\lambda||\kappa| v s_{\phi_y}s_\beta\,,\quad 
\lambda_{144}=\fr{c_\beta M_Z^2}{v} \,,\quad 
 \quad \lambda_{145}=0\,, \nonumber \\
\lambda_{146}&=0\,,\quad\lambda_{155}=-\fr{c_\beta M_Z^2}{v}+ |\lambda|^2 c_\beta v \,, \quad \lambda_{156}= \fr{|A_\lambda| |\lambda|c_{\phi_x}}{\sqrt{2}} - |\lambda||\kappa| v_s c_{\phi_y}\,, 
\nonumber\\
 \lambda_{166}&= |\lambda|^2 c_\beta v + |\kappa| |\lambda| v s_\beta  c_{\phi_y}\,,\quad
\lambda_{222}= \fr{3M_Z^2 s_\beta}{v}\,, \quad \lambda_{223}=|\lambda|^2 v_s\,, \quad\lambda_{224}= 0\,,\quad \lambda_{225}=0 \,,
\nonumber\\
\lambda_{226}&=0\,, \quad \lambda_{233}= -|\lambda||\kappa| v c_\beta  c_{\phi_y} + |\lambda|^2 s_\beta v\,,\quad \lambda_{234}=\fr12 |\kappa||\lambda| v_s s_{\phi_y} \,, \quad  \lambda_{235}= 0\,, 
\nonumber\\
\lambda_{236}&=- |\kappa||\lambda| vc_\beta s_{\phi_y}\,,\quad 
\lambda_{244}=-\fr{M_Z^2 s_\beta}{v} + |\lambda|^2 v s_\beta \,, \quad \lambda_{245}=0\,, 
\nonumber\\ \lambda_{246}&=\fr{|A_\lambda| |\lambda|c_{\phi_x}}{\sqrt{2}} - |\lambda||\kappa| v_s c_{\phi_y}\,,\quad
\lambda_{255}= \fr{M_Z^2 s_\beta}{v}\,, \quad \lambda_{256}=0\,, 
\nonumber\\
\quad \lambda_{266}&= |\lambda||\kappa| v c_\beta c_{\phi_y} + |\lambda|^2 s_\beta v \,,\quad
\lambda_{333}=  6|\kappa|^2 v_s +\sqrt{2} |A_\kappa||\kappa| c_{\phi_z}  \,, \quad \lambda_{334}=|\kappa||\lambda| v s_\beta s_{\phi_y} \,, \nonumber
\end{align}
\begin{align}
\lambda_{335}&=|\kappa||\lambda|v c_\beta  s_{\phi_y} \,,\quad
\lambda_{336}=\fr{3|\lambda||\kappa|s_\beta c_\beta s_{\phi_y}
  v^2}{v_s} \,, \quad\lambda_{344}=|\lambda|^2 v_s \,, \nonumber \\
\lambda_{345}&=\fr{|A_\lambda||\lambda| c_{\phi_x}}{\sqrt{2}}|\kappa||\lambda| v_s c_{\phi_y}\,,\quad
\lambda_{346}=-|\kappa||\lambda| v  s_\beta c_{\phi_y}  \,, \quad \lambda_{355}=|\lambda|^2 v_s\,, 
\nonumber \\
\lambda_{356}&=-|\kappa||\lambda|v c_\beta  c_{\phi_y} \,,
\lambda_{366}= -\sqrt{2}|A_\kappa||\kappa|c_{\phi_z} +
2|\kappa|^2v_s\,, \quad \lambda_{444}=0\,, \quad \lambda_{445}=0\,, 
\nonumber \\
\lambda_{446}&=0 \,, \quad \lambda_{455}=0\,, \quad \lambda_{456}= \fr32|\kappa||\lambda|v_s s_{\phi_y}\,,\quad \lambda_{466}= -|\kappa||\lambda|v s_\beta s_{\phi_y} \,, 
\nonumber\\
 \lambda_{555}&=0\,, \quad \lambda_{556}=0\,,\quad 
\lambda_{566}= -|\kappa||\lambda|v c_\beta s_{\phi_y}\,, \quad \lambda_{666}=-\fr{3 |\kappa||\lambda| c_\beta s_\beta s_{\phi_y}v^2}{v_s}\,.
\end{align}

\section{The order ${\cal O}(\al_t)$ corrections to the trilinear
  Higgs self-couplings \label{sec:one-loop-trilinear}}
The order ${\cal O}(\al_t)$ one-loop corrections introduced in
Eq.~(\ref{eq:decomposition}), $\Delta^{(1)}\lambda^{\text{UR}}_{\phi_i \phi_j
  \phi_k} \equiv \Delta^{(1)}_{ijk}$ 
($i,j,k=1,...,6$), with the same correspondences as introduced in
Appendix~\ref{sec:tree-leve-trilinear}, can be cast into the form
\begin{align}
\Delta^{(1)}_{ijk}& =-2 C_F m_t y_t^3  \left[F_{1x}\left( (h_i^t)^* h_j^t h_k^t+ h_i^t
   (h_j^t)^* h_k^t+ h_i^t h_j^t (h_k^t)^*\right)+h_i^t
 h_j^th_k^t+\text{c.c.}\right]\\ \non 
&-C_F y_t^3\bigg[\left(-F_{3 x}  y_{h_i\tilde{t}_2\tilde{t}_1} y_{h_j\tilde{t}_1 \tilde{t}_2}
   y_{h_k\tilde{t}_1 \tilde{t}_1}+F_{2 x}  y_{h_i\tilde{t}_2 \tilde{t}_2}
   y_{h_j\tilde{t}_2 \tilde{t}_1} y_{h_k\tilde{t}_1 \tilde{t}_2}+ \text{Permutation}[i,j,k]\right)\\ \non
&-\frac{
   y_{h_i\tilde{t}_1 \tilde{t}_1} y_{h_j\tilde{t}_1 \tilde{t}_1} y_{h_k\tilde{t}_1
   \tilde{t}_1}}{m_{\tilde{t}_1}^2}-\frac{ y_{h_i\tilde{t}_2 \tilde{t}_2}
   y_{h_j\tilde{t}_2 \tilde{t}_2} y_{h_k\tilde{t}_2 \tilde{t}_2}}{m_{\tilde{t}_2}^2}\bigg]\\ \non
&-C_F y_t^2\bigg[ F_{4 x}  \left(y_{h_k\tilde{t}_2 \tilde{t}_1} y_{h_ih_j\tilde{t}_1\tilde{t}_2}+
   y_{h_k\tilde{t}_1 \tilde{t}_2} y_{h_ih_j\tilde{t}_2\tilde{t}_1}\right)
-\log\frac{m_{\tilde{t}_1}^2}{Q^2} y_{h_k\tilde{t}_1
   \tilde{t}_1} y_{h_ih_j\tilde{t}_1\tilde{t}_1}\\ \non
&-\log\frac{m_{\tilde{t}_2}^2}{Q^2} y_{h_k\tilde{t}_2
   \tilde{t}_2} y_{h_ih_j\tilde{t}_2\tilde{t}_2}+(k\leftrightarrow i)+(k\leftrightarrow j)\bigg]
\end{align}
where
\begin{align}
C_F&=\frac{3}{16 \pi^2}\,,\quad y_t= \fr{\sqrt{2}m_t}{vs_\beta}\,,\quad  F_{1x}= 2
\log\fr{m_t^2}{Q^2} + 1\,,\quad F_{2 x}=\fr{m_{\tilde{t}_1}^2 -
  m_{\tilde{t}_2}^2 -
  m_{\tilde{t}_1}^2\log\fr{m_{\tilde{t}_1}^2}{m_{\tilde{t}_2}^2}
}{(m_{\tilde{t}_1}^2-m_{\tilde{t}_2}^2)^2}\,,
\end{align}
\begin{align}
 F_{3 x}&=\fr{m_{\tilde{t}_1}^2 - m_{\tilde{t}_2}^2 - m_{\tilde{t}_2}^2\log\fr{m_{\tilde{t}_1}^2}{m_{\tilde{t}_2}^2} }{(m_{\tilde{t}_1}^2-m_{\tilde{t}_2}^2)^2}\,,\quad F_{4 x}=\fr{m_{\tilde{t}_1}^2 - m_{\tilde{t}_2}^2 - m_{\tilde{t}_1}^2\log\fr{m_{\tilde{t}_1}^2}{Q^2}+m_{\tilde{t}_2}^2\log\fr{m_{\tilde{t}_2}^2}{Q^2} }{(m_{\tilde{t}_1}^2-m_{\tilde{t}_2}^2)}\,,
\end{align}
and the non-vanishing couplings read
\begin{align}
h_2^t&=\fr{1}{\sqrt{2}}\,,\quad h_5^t=\fr{i}{\sqrt{2}}\,,\quad  y_{h_1\tilde{t}_n \tilde{t}_m}=-\frac{1}{\sqrt{2}} \mu ^* \mathcal{U}_{\tilde{t}_{n1}}^*
   \mathcal{U}_{\tilde{t}_{m2}}-\frac{1}{\sqrt{2}} \mu 
   \mathcal{U}_{\tilde{t}_{n2}}^* \mathcal{U}_{\tilde{t}_{m1}}\,,\\ \non
y_{h_2\tilde{t}_n \tilde{t}_m}&=\frac{A_t e^{i \varphi_u}
   \mathcal{U}_{\tilde{t}_{m2}}
   \mathcal{U}_{\tilde{t}_{n1}}^*}{\sqrt{2}}+\frac{A_t^*
   e^{-i \varphi_u} \mathcal{U}_{\tilde{t}_{m1}}
   \mathcal{U}_{\tilde{t}_{n2}}^*}{\sqrt{2}}+\sqrt{2} m_t
   \mathcal{U}_{\tilde{t}_{m1}} \mathcal{U}_{\tilde{t}_{n1}}^*+\sqrt{2}
   m_t \mathcal{U}_{\tilde{t}_{m2}} \mathcal{U}_{\tilde{t}_{n2}}^*\,, \\\non
 y_{h_3\tilde{t}_n \tilde{t}_m}&= -\frac{\lambda ^* c_{\beta }v e^{-i
   \varphi_s} \mathcal{U}_{\tilde{t}_{m,2}}
   \mathcal{U}_{\tilde{t}_{n,1}}^*}{2}-\frac{\lambda 
   c_{\beta } v e^{i \varphi_s} \mathcal{U}_{\tilde{t}_{m,1}}
   \mathcal{U}_{\tilde{t}_{n,2}}^*}{2}\,, \\
\non
 y_{h_4\tilde{t}_n \tilde{t}_m}&=\frac{1}{\sqrt{2}} i \mu ^*
   \mathcal{U}_{\tilde{t}_{m2}}
   \mathcal{U}_{\tilde{t}_{n1}}^*-\frac{1}{\sqrt{2}} i \mu 
   \mathcal{U}_{\tilde{t}_{m1}} \mathcal{U}_{\tilde{t}_{n2}}^* \,, 
\end{align}
\begin{align}
\non
 y_{h_5\tilde{t}_n \tilde{t}_m}&=  \frac{i A_t e^{i \varphi_u}
   \mathcal{U}_{\tilde{t}_{m2}}
  \mathcal{U}_{\tilde{t}_{n1}}^*}{\sqrt{2}}-\frac{i A_t^*
   e^{-i \varphi_u} \mathcal{U}_{\tilde{t}_{m1}}
   \mathcal{U}_{\tilde{t}_{n2}}^*}{\sqrt{2}}\,, \\ \non
 y_{h_6\tilde{t}_n \tilde{t}_m}&= \frac{i \lambda ^* c_{\beta } v e^{-i
   \varphi_s} \mathcal{U}_{\tilde{t}_{m2}}
   \mathcal{U}_{\tilde{t}_{n1}}^*}{2}-\frac{i \lambda 
   c_{\beta } v e^{i \varphi_s} \mathcal{U}_{\tilde{t}_{m1}}
   \mathcal{U}_{\tilde{t}_{n2}}^*}{2}\,, \\
\non
y_{h_1h_3\tilde{t}_n\tilde{t}_m}&=-\frac{1}{2} \lambda ^* e^{-i \varphi_s} \mathcal{U}_{\tilde{t}_{m2}}
   \mathcal{U}_{\tilde{t}_{n1}}^*-\frac{1}{2} \lambda  e^{i \varphi_s}
   \mathcal{U}_{\tilde{t}_{m1}} \mathcal{U}_{\tilde{t}_{n2}}^* \,,\\ \non
y_{h_1h_6\tilde{t}_n\tilde{t}_m}&= \frac{1}{2} i \lambda ^* e^{-i
  \varphi_s} \mathcal{U}_{\tilde{t}_{m2}} 
   \mathcal{U}_{\tilde{t}_{n1}}^*-\frac{1}{2} i \lambda  e^{i \varphi_s}
   \mathcal{U}_{\tilde{t}_{m1}} \mathcal{U}_{\tilde{t}_{n2}}^* \,,\\ \non
y_{h_2h_2\tilde{t}_n\tilde{t}_m}&=y_t \mathcal{U}_{\tilde{t}_{m1}}
  \mathcal{U}_{\tilde{t}_{n1}}^*+y_t\mathcal{U}_{\tilde{t}_{m2}}
   \mathcal{U}_{\tilde{t}_{n2}}^* \,,\\ \non
y_{h_3h_4\tilde{t}_n\tilde{t}_m}&=\frac{1}{2} i \lambda ^* e^{-i \varphi_s} \mathcal{U}_{\tilde{t}_{m2}}
   \mathcal{U}_{\tilde{t}_{n1}}^*-\frac{1}{2} i \lambda  e^{i \varphi_s}
   \mathcal{U}_{\tilde{t}_{m1}} \mathcal{U}_{\tilde{t}_{n2}}^*\,,\\ \non
y_{h_4h_6\tilde{t}_n\tilde{t}_m}&= \frac{1}{2} \lambda ^* e^{-i \varphi_s}
   \mathcal{U}_{\tilde{t}_{m2}}
   \mathcal{U}_{\tilde{t}_{n1}}^*+\frac{1}{2} \lambda  e^{i \varphi_s}
   \mathcal{U}_{\tilde{t}_{m1}} \mathcal{U}_{\tilde{t}_{n2}}^* \,,\\ \non
y_{h_5h_5\tilde{t}_n\tilde{t}_m}&=y_t\mathcal{U}_{\tilde{t}_{m1}}
  \mathcal{U}_{\tilde{t}_{n1}}^*+y_t\mathcal{U}_{\tilde{t}_{m2}}
  \mathcal{U}_{\tilde{t}_{n2}}^* \;.
\end{align}

\section{The trilinear Higgs self-coupling
  counterterms \label{sec:trilinearCT}}
Here we summarize the one- and two-loop ($l=1,2$) non-vanishing counterterms $\Delta^{(l)}
\lambda^{\text{CT}}_{ijk}$ that arise in the computation of the
loop-corrected Higgs self-couplings $\lambda_{ijk}$
($i,j,k=1,...,6)$. They are given in the interaction basis, and read
in terms of the various tadpole, mass, wave function and parameter counterterms as
\begin{align}
\Del_{112}&=2 v \asbe \asla \deltalla + v \acbe^3 \asla^2 \deltalTB + 
 \asbe \asla^2 \deltalv + \fr12 v \asbe \asla^2 \delzhu\,, \\ \non
\Del_{113}&=2 v_s \asla \deltalla  \,, \\ \non
 \Del_{122}&= 2 v \acbe \asla \deltalla - v \acbe^2  \asbe \asla^2  \deltalTB + 
 \acbe \asla^2 \deltalv + v \acbe \asla^2 \delzhu, \\ \non
\Del_{123}&=\left(-\fr12 v_s \acphi \aska - \fr{v^2 \acbe \asbe \asla}{v_s}\right) \deltalla 
- \fr{ c_{2\beta} \acbe^2 (2 \aMHp + v^2 \asla^2) \deltalTB}{2 v_s} \\ \non
&+ \fr{ \asbe^3 \deltalthd + \acbe^3 \deltalthu}{v v_s} - \fr{\acbe \asbe \dMHpsq} {v_s}- \fr{
 v \acbe \asbe \asla^2 \deltalv}{v_s} \\ \non
 &- \fr{(s_{2\beta} \aMHp + 
    v_s^2 \acphi \aska \asla + v^2 \acbe \asbe \asla^2) \delzhu}{
 4 v_s}  \,,\\ \non
\Del_{126}&= -\fr32 v_s\asphi \aska  \deltalla  - 
 \fr34 v_s \asphi \aska \asla \delzhu  + \fr{\deltaltad}{v v_s \asbe}, \\ \non
\Del_{133}&=v (-\acphi \asbe \aska + 2 \acbe \asla) \deltalla - 
 v \acbe^2 \asla (\acbe \acphi \aska + \asbe \asla) \deltalTB \\ \non
 &+ 
 \asla (-\acphi \asbe \aska + \acbe \asla) \deltalv, \\
\non
  \Del_{135}&=\fr12 v_s \aska \asphi \deltalla  + 
 \fr14 v_s \aska \asla \asphi \delzhu + \fr\deltaltad{v v_s \asbe}, \\  \non
  \Del_{136}&=-v \asbe \asphi \aska  \deltalla - v \acbe^3 \asphi \aska \asla  \deltalTB - 
 \asbe \asphi\aska \asla  \deltalv, \\  \non
    \Del_{155}&=\Del_{122},\qquad
  \Del_{156}=-\Del_{123}, \\
\non
  \Del_{166}&=v (\acphi \asbe \aska + 2 \acbe \asla) \deltalla + 
 v \acbe^2 \asla (\acbe \acphi \aska - \asbe \asla) \deltalTB \\\non
 &+ 
 \asla (\acphi \asbe \aska + \acbe \asla) \deltalv, 
\end{align}
\begin{align}
\non
\Del_{223}&=2 v_s \asla \deltalla + v_s \asla^2 \delzhu, \\
\non
\Del_{233}&=(-v \acbe \acphi \aska + 2 v \asbe \asla) \deltalla + 
 v \acbe^2 \asla (\acphi \asbe \aska + \acbe \asla) \deltalTB \\\non
 &+ 
 \asla (-\acbe \acphi \aska + \asbe \asla) \deltalv + 
 \fr12 v \asla (-\acbe \acphi \aska + \asbe \asla) \delzhu, \\
\non
\Del_{234}&=\Del_{135}, \\  \non
  \Del_{236}&=-v \acbe \aska \asphi \deltalla + 
 v \acbe^2 \asbe \aska \asla \asphi \deltalTB - 
 \acbe \aska \asla \asphi \deltalv - \fr12 v \acbe \aska \asla \asphi \delzhu, \\  \non
   \Del_{244}&=\Del_{112}, \\  \non
   \Del_{246}&=\left(-\fr32 v_s \acphi \aska + \fr{v^2 \acbe \asbe \asla}{v_s} \right)\deltalla + \fr{
 c_{2\beta} \acbe^2 (2 \aMHp + v^2 \asla^2) \deltalTB}{2 v_s}\\ \non
 &-
 \fr{\asbe^3 \deltalthd + \acbe^3 \deltalthu}{v v_s} + \fr{
 v \acbe \asbe \asla^2 \deltalv}{v_s}+ \fr{\acbe \asbe \dMHpsq}{v_s}\\\non
 &+ \fr{(s_{2\beta} \aMHp  - 
    3 v_s^2 \acphi \aska \asla + v^2 \acbe \asbe \asla^2) \delzhu}{
 4 v_s} , \\ \non 
    \Del_{266}&=v (\acbe \acphi \aska + 2 \asbe \asla) \deltalla + 
 v \acbe^2 \asla (-\acphi \asbe \aska + \acbe \asla) \deltalTB \\ \non&+ 
 \asla (\acbe \acphi \aska + \asbe \asla) \deltalv + 
 \fr12 v \asla (\acbe \acphi \aska + \asbe \asla) \delzhu, \\  \non
      \Del_{333}&=\fr{3 v^2 \acbe \asbe \asphi t_{\phi_z} \aska  \deltalla }{v_s} + \fr{
 3 v^2 \acbe^2 c_{2\beta}\asphi  t_{\phi_z}\aska \asla  \deltalTB }{v_s}  + \fr{
 6 v \acbe \asbe \asphi  t_{\phi_z} \aska \asla \deltalv }{v_s}, \\ \non
        \Del_{334}&=v \asbe\asphi\aska  \deltalla + v \acbe^3 \asphi\aska \asla  \deltalTB + 
 \asbe \asphi \aska \asla  \deltalv, \\  \non
  \Del_{335}&=v \acbe \asphi \aska \deltalla - 
 v \acbe^2 \asbe \asphi \aska \asla  \deltalTB + 
 \acbe \asphi\aska \asla  \deltalv + \fr12 v \acbe \asphi \aska \asla  \delzhu, \\  \non
    \Del_{336}&= \fr{3 v^2 \acbe \asbe \asphi \aska  \deltalla}{v_s} + \fr{
 3 v^2 \acbe^2 c_{2\beta} \asphi \aska \asla  \deltalTB}{v_s} + \fr{
 6 v \acbe \asbe \asphi \aska \asla  \deltalv}{v_s}, \\  \non
      \Del_{344}&= \Del_{113}, \qquad
    \Del_{345}=-\Del_{123}, \qquad
    \Del_{346}=\fr\acphi\asphi\Del_{136}, \\  \non
    \Del_{355}&=\Del_{223}, \qquad
    \Del_{356}=\fr\acphi\asphi\Del_{335}, \qquad
    \Del_{366}=-\Del_{333}, \\  \non
    \Del_{456}&=-\Del_{126}, \quad
    \Del_{466}=-\Del_{334}, \quad
    \Del_{566}=-\Del_{335}, \quad
    \Del_{666}=-\Del_{336}. \\  \non
\end{align}

\section{Computation of $\overline{\mbox{DR}}$ parameters at
    different scales \label{app:trick}}
The values of the $\overline{\mbox{DR}}$ parameters at the scale $\mu$
are obtained by renormalization group running from the starting scale
$\mu_0$ to the scale $\mu$. If the scales are not too far apart an
approximate result can be obtained by exploiting the relation between
OS and $\overline{\mbox{DR}}$ parameters $p$ at the scale $\mu$,
\beq
p^{\text{OS}} + \delta p^{\text{OS}} (\mu) = p^{\overline{\mbox{DR}}} (\mu)
+ \delta p^{\overline{\mbox{DR}}} (\mu) \;.
\eeq
Here $\delta p^{\text{OS}}$ and $\delta p^{\overline{\mbox{DR}}}$
denote the OS and ${\overline{\mbox{DR}}}$ counterterm,
respectively. The scale dependence in the $\overline{\mbox{DR}}$
counterterm, which purely subtracts the UV divergences, enters through
the scale dependence of the parameters. As has been shown in
\cite{Muhlleitner:2014vsa} the only $\overline{\mbox{DR}}$ parameters
that receive two-loop counterterms at order ${\cal O} (\alpha_t
\alpha_s)$ are $\tan\beta$ and $\lambda$, and this 
arises due to the non-vanishing wave function renormalization
counterterm for $H_u$. We exemplify for $\tan\beta$ how to obtain the
relation between the $\overline{\mbox{DR}}$ renormalized $\tan\beta$'s
at two different scales $\mu_1$ and $\mu_2$. We denote by
$\tan\beta^{\text{pure}\overline{\mbox{DR}}}$ the $\tan\beta$ defined
through the $\overline{\mbox{DR}}$ condition with the top/stop sector
renormalized $\overline{\mbox{DR}}$. Analogously,
$\tan\beta^{\text{pureOS}}$ is understood to be the OS $\tan\beta$ and
the top/stop sector renormalized in the OS scheme. The relation
between these two definitions of $\tan\beta$ is given by,
\beq
\tan\beta^{\text{pureOS}} + \delta^{(1)} \tan\beta^{\text{pureOS}} +
\delta^{(2)} \tan\beta^{\text{pureOS}} &=& \label{eq:relationOSDR}  \\ &&
\hspace*{-3cm}
\tan\beta^{\text{pure}\overline{\mbox{DR}}} + \delta^{(1)}
\tan\beta^{\text{pure}\overline{\mbox{DR}}} + \delta^{(2)}
\tan\beta^{\text{pure}\overline{\mbox{DR}}} \,, \nonumber
\eeq
where again the superscripts $(1)$ and $(2)$ refer to the one- and
two-loop counterterm, respectively. The one- and two-loop counterterms in
the pure OS scheme can be expanded in terms of $\epsilon$ as
\beq
\delta^{(1)} \tan\beta^{\text{pureOS}}\hspace*{-0.3cm} &=&
\hspace*{-0.3cm}\mu^{2\epsilon} \left( 
  \frac{a_1(m_t^{\text{OS}})}{\epsilon} + f_1 (m_t^{\text{OS}})
\right) \\
\delta^{(2)} \tan\beta^{\text{pureOS}}\hspace*{-0.3cm} &=&
\hspace*{-0.3cm}\mu^{4\epsilon} \left( 
  \frac{b_2(m_t^{\text{OS}},\alpha_s^{\DRb}(\mu))}{\epsilon^2} + 
  \frac{a_2(m_t^{\text{OS}},\alpha_s^{\DRb}(\mu))}{\epsilon} + f_2
  (m_t^{\text{OS}},\al_s^{\DRb}(\mu))  
\right) 
\eeq
where the functions $a_1$ and $f_1$ do not depend on the
renormalization scale $\mu$ while $a_2$, $b_2$ and $f_2$ implicitly depend
on $\mu$ through their dependence on $\al_s^{\DRb}(\mu)$. Note that
these expansions can only be 
applied in the OS scheme of the top/stop sector in the context of our
calculation. In the limit $\epsilon \to 0$ these equations read
\beq
\delta^{(1)} \tan\beta^{\text{pureOS}} &=&
\frac{a_1(m_t^{\text{OS}})}{\epsilon} + a_1(m_t^{\text{OS}}) \ln \mu^2
+ f_1 (m_t^{\text{OS}})  \label{eq:1lexpansion}\\
\delta^{(2)} \tan\beta^{\text{pureOS}} &=& 
  \frac{b_2(m_t^{\text{OS}},\al_s^{\DRb}(\mu))}{\epsilon^2} +
  \frac{a_2(m_t^{\text{OS}},\al_s^{\DRb}(\mu)) 
  + 2\, b_2(m_t^{\text{OS}},\al_s^{\DRb}(\mu)) \ln \mu^2}{\epsilon}
\nonumber \\
 && + 2\, a_2(m_t^{\text{OS}},\al_s^{\DRb}(\mu)) \ln \mu^2 
  + 2\, b_2(m_t^{\text{OS}},\al_s^{\DRb}(\mu)) \ln^2
  \mu^2 \nonumber \\
&& + f_2 (m_t^{\text{OS}},\al_s^{\DRb}(\mu)) \;. \label{eq:2lexpansion}
\eeq
The one- and two-loop counterterms in the pure $\DRb$ scheme are
\bea
 \deltaone \tan\beta^{\text{pure}\DRb} &=& \fr{a_1(m_t^{\DRb}(\mu))}{\epsilon} ,\label{eq:1lexpansionDR} \\
\deltatwo \tan\beta^{\text{pure}\DRb} &=&
\fr{b_2(m_t^{\DRb}(\mu),\al_s^{\DRb}(\mu))}{\epsilon^2}+\fr{c_2(m_t^{\DRb}(\mu),\al_s^{\DRb}(\mu))}{\epsilon} \,. \label{eq:2lexpansionDR} 
\eea
Replacing Eqs.~(\ref{eq:1lexpansion}), (\ref{eq:2lexpansion}),
(\ref{eq:1lexpansionDR}) and (\ref{eq:2lexpansionDR}) into
Eq.~(\ref{eq:relationOSDR}) one gets the relation of the pure
$\overline{\mbox{DR}}$ renormalized $\tan\beta$'s at the scales $\mu_1$
and $\mu_2$. Taking into account the relation
\beq
m_t^{\overline{\mbox{DR}}} = m_t^{\text{OS}} + (\delta
m_t)_{\text{fin}} \,,
\eeq
where $(\delta m_t)_{\text{fin}}$ denotes the finite part of the OS
counterterm, all terms proportional to the poles in $\epsilon$ cancel
at the considered order, and we are left with
\beq
\tan\beta^{\text{pure}\DRb}(\mu_1)&\hspace*{-0.2cm}-&\hspace*{-0.2cm} \tan\beta^{\text{pure}\DRb}(\mu_2)=
  a_1(m_t^\OS) \ln \fr{\mu_1^2}{\mu_2^2} \nonumber \\
&& + 2\, a_2(m_t^\OS,\al_s^{\DRb}(\mu_1)) \ln \mu_1^2 
- 2\, a_2(m_t^\OS,\al_s^{\DRb}(\mu_2)) \ln \mu_2^2 \nonumber \\
&& + 2\, b_2(m_t^\OS,\al_s^{\DRb}(\mu_1)) \ln^2\mu_1^2 
- 2\, b_2(m_t^\OS,\al_s^{\DRb}(\mu_2)) \ln^2 \mu_2^2 \,. 
\eeq
For the parameters $p$ that are renormalized at one-loop order only,
this relation simplifies to
\beq
p^{\text{pure}\DRb}(\mu_1) - p^{\text{pure}\DRb}(\mu_2) = a_1 \ln
\frac{\mu_1^2}{\mu_2^2} \,.
\eeq

\end{appendix}

\bibliographystyle{h-physrev}


\end{document}